\documentclass[twoside,12pt]{article}
\usepackage{epsfig}

\newcommand{\be}{\begin{equation}}
\newcommand{\ee}{\end{equation}}
\newcommand{\bea}{\begin{eqnarray}}
\newcommand{\eea}{\end{eqnarray}}

\begin{document}

\title{Temperature and momentum dependence of single-particle properties
in hot asymmetric nuclear matter }

\author{Ch.C. Moustakidis\\
$^{}$ Department of Theoretical Physics, Aristotle University of
Thessaloniki, \\ 54124 Thessaloniki, Greece }

\maketitle

\begin{abstract}
We have studied the effects of momentum dependent interactions on
the single-particle properties of hot asymmetric nuclear matter.
In particular, the single-particle potential of protons and
neutrons as well as the symmetry potential have been studied
within a self-consistent model using a momentum dependent
effective interaction. In addition, the isospin splitting of the
effective mass has been derived from the above model. In each case
temperature effects have been included and analyzed. The role of
the specific parametrization of the effective interaction used in
the present work has been investigated. It has been concluded that
the behavior of the symmetry potential depends strongly on the
parametrization of the interaction part of the energy density and
the momentum dependence of the regulator function. The effects of
the parametrization have been found to be less pronounced on the
isospin mass splitting.
\\
\\
PACS number(s): 21.65.+f, 21.30.Fe, 24.10.Pa, 26.60.+c
\end{abstract}

\section{Introduction}
One of the most interesting problems in nuclear physics is the
isovector dependence on  nuclear force, which can be found in
nuclear symmetry energy, the isovector optical potential and
neutron-proton effective mass splitting. The isovector feature of
nuclear forces is crucial in order to gain a good understanding of
neutron stars and exotic nuclear collisions produced at
radioactive beam facilities and to describe the structure of
exotic nuclei. This interest in the isospin dependence of nuclear
forces is of recent date because data for neutron-rich nuclei were
rather scarce in the past. The forthcoming new generation of
radioactive beam facilities, such as the future GSI facility FAIR,
the Rare Isotope Accelerator planned in the USA  and the SPIRAL2
at GANIL, will produce huge amounts of new data for neutron-rich
nuclei. Up to now, the isovector dependence of nuclear force has
been investigated in  heavy ion collisions experiments. The
advantage of such kinds of reactions is that they allow   testing
of nuclear forces at supranormal densities since in intermediate
energy compressions of two to three times nuclear saturation
density is reached. Nevertheless, the asymmetry of the colliding
systems is moderate and therefore the isospin effects on the
corresponding observables are moderate as well.

From a theoretical point of view, the predictions for the isospin
dependence of nuclear interaction are very different. In general,
both microscopic and effective interactions have been extensively
used in order to gain knowledge about the nuclear matter
properties in conditions far from equilibrium (hot nuclear matter,
high density behavior and so on). It is appropriate therefore, in
every case to incorporate temperature and momentum dependence in
order to have a richer interaction and as a consequence  be able
to produce a more thorough description of
 nuclear matter properties.. Specifically, the determination of the nuclear
symmetry energy (NSE) based on microscopic and/or phenomenological
approaches is of great interest to nuclear physics and nuclear
astrophysics. For instance, it is important for the study of the
structure and reactions of neutron-rich nuclei, the Type II
supernova explosions,  neutron-star mergers and  the stability of
neutron stars. So far, the main part of the calculations
concerning the density dependence of the SE is related to cold
nuclear matter ($T=0$). However, recently there has been an
increasing interest in the study of the properties of asymmetric
nuclear matter, including NSE, and  the properties of neutron
stars at finite temperature
\cite{Prakash-97,Kupper-74,Lamb-81,Muller-95,Danielewicz-02,Li-06,Xu-07-1,Xu-07-2,Xu-07-3,Bao-Li-06,Mishra-93,
Ccernai-92,Zuo-03,Lee-01,Das-07,Zuo-06,Huber-98,Das-93,Bandy-90,Jena-04,Moustakidis-07}

In our previous work \cite{Moustakidis-07} we studied the effects
of finite temperature on NSE  and we also found the appropriate
relations describing that effect. We especially focused on the
interaction part of the NSE, which  so far has received little
theoretical attention concerning its dependence on temperature. We
applied a momentum dependent effective interaction model. This
model, afterwards called BGBD (Bombaci, Gale, Bertsch, Das
Gupta),has been introduced by Gale et al.
\cite{Gale-87,Gale-90,Bertsch-88,Prakash-88-1} to examine the
influence of momentum-dependent interactions on the collective
flow of heavy ion collisions. Over the years the model has been
extensively applied to study not only  the heavy ion collisions,
but also the properties of nuclear matter by proper modification
\cite{Prakash-97,Bombaci-01,Das-03,Bao-Li-04,Chen-05}. A review
analysis of the present model is presented in Refs.
\cite{Prakash-97,Bertsch-88}.

In the present work, we have studied the momentum and temperature
dependence of the mean field properties of asymmetric nuclear
matter
\cite{Hodgson-94,Jeukenne-76,Baran-05,Jeukenne-75,Wiringa-88,Daniel-00,Bombaci-91,BLi-04,
Zuo-99,Zuo-05,Zuo-07,Sammarruca-05,Rizzo-04,Rizzo-05,Toro-06,
Dalen-05,Bao05,Bao04,Behera-98,Behera-05,LEChen-07,LWChen-05,HuagLi-06}.
Efforts up to now  have been devoted mostly to studying the
properties of cold nuclear matter. In contrast, the motivation for
the present work is to study the properties of hot asymmetric
nuclear matter, especially the temperature dependence of the
nuclear symmetry energy and the single particle properties of
nuclear matter.

We have employed a model with the characteristic property that the
interacting part of the energy density is momentum dependent.
Thus, the single particle potentials of protons, neutrons, as well
as the  symmetry potential, are also momentum dependent. The
isovector part of the optical potential, i.e. the symmetry
potential, describes the difference between the neutron and proton
single particle potentials in neutron rich matter. The symmetry
potential is one of the basic inputs to the transport models for
the collisions of radioactive nuclei \cite{Hodgson-94}. In
addition, due to the momentum dependence, the temperature is
expected to affect not only the kinetic part, but also the
interacting part of the energy density. This is important in the
sense that the density dependence of the nuclear symmetry energy,
influenced by temperature, has a powerful effect on the values of
the proton fraction and consequently  the composition of hot
$\beta$-stable nuclear matter, with extensive applications in
heavy-ion collisions and Nuclear Astrophysics.

The effective mass (EM) is one of the most fundamental properties
characterizing the propagation of a nucleon in a nuclear medium.
Knowledge about nucleon EM in neutron rich matter is crucial to
fully understand several properties of neutron stars. EM is
determined by the derivative of the single particle potential with
respect to the momenta $k$ for $k=k_F$. Thus, the trend of the EM
is directly connected to the momentum dependence of the
corresponding single particle potential. Furthermore, the isospin
splitting of the effective mass i.e. the difference between the
neutron and proton effective masses is derived from the above
model. The present study is a contribution to the theoretical
study of the neutron-proton effective mass splitting, a problem
which is still highly controversial within different approaches
and/or using different nuclear effective interactions
\cite{Xu-07-2}. The above analysis indicates the necessity to
apply a momentum dependent interacting model to study the single
particle properties of  hot nuclear matter.


This work is a continuation of recent papers
\cite{BLi-04,Rizzo-04}, where the authors have employed a
phenomenological non-relativistic effective interaction first
introduced in Refs. \cite{Prakash-97,Bombaci-01}, with  suitable
modification concerning the isovector part of the interaction.
Here, we have applied a more generalized expression of the
interaction part of the energy density, with a richer
parametrization and additional parameters introduced to maintain
causality \cite{Prakash-97}. We have concentrated on a systematic
study of the effect of the parametrization of the effective
interaction on the mean field properties of  hot asymmetric
nuclear matter.

The plan of the paper is as follows. In Sec.~II the model and the
related formulae are discussed and analyzed. Results are reported
and discussed in Sec.~III, while Sec.~IV contains a summary.

\section{The  model}
The schematic potential model used in the present work, is
designed  to reproduce the results of  microscopic calculations of
both nuclear and neutron-rich matter at zero temperature and can
be extended to finite temperature \cite{Prakash-97,Bombaci-01}.
The energy density of the asymmetric nuclear matter (ANM) is given
by the relation
\begin{equation}
\epsilon(n_n,n_p,T)=\epsilon_{kin}^{n}(n_n,T)+\epsilon_{kin}^{p}(n_p,T)+
V_{int}(n_n,n_p,T), \label{E-D-1}
\end{equation}
where $n_n$ ($n_p$) is the neutron (proton) density and the total
baryon density is $n=n_n+n_p$. The contribution of the kinetic
parts is
\begin{equation}
\epsilon_{kin}^n(n_n,T)+\epsilon_{kin}^p(n_p,T)=2 \int \frac{d^3
k}{(2 \pi)^3}\frac{\hbar^2 k^2}{2m}
\left(f_n(n_n,k,T)+f_p(n_p,k,T) \right), \label{E-K-D-1}
\end{equation}
where $f_{\tau}$, (for $\tau=n,p$) is the Fermi-Dirac distribution
function with the form
\begin{equation}
f_{\tau}(n_{\tau},k,T)=\left[1+\exp\left(\frac{e_{\tau}(n_{\tau},k,T)-\mu_{\tau}(n_{\tau},T)}{T}\right)
\right]^{-1}. \label{FD-1}
\end{equation}
This distribution is inserted into the following integral in order
to evaluate the nucleon density $n_{\tau}$,
\begin{equation}
n_{\tau}=2 \int \frac{d^3k}{(2\pi)^3}f_{\tau}(n_{\tau},k,T).
\label{D-1}
\end{equation}
In Eq. (\ref{FD-1}), $e_{\tau}(n_{\tau},k,T)$ is the single
particle energy (SPE) having the form
\begin{equation}
e_{\tau}(n_{\tau},k,T)=\frac{\hbar^2k^2}{2m}+U_{\tau}(n_{\tau},k,T).
\label{esp-1}
\end{equation}
$\mu_{\tau}(n_{\tau},T)$ stands for the chemical potential of each
species, while the single particle potential
$U_{\tau}(n_{\tau},k,T)$ is obtained by the functional derivative
of the interaction part of the energy density with respect to the
distribution function $f_{\tau}$.

Including the effect of finite-range forces between nucleons to
avoid acausal behavior at high densities, the potential
contribution is parameterized as follows
\begin{eqnarray}
V_{int}(n_n,n_p,T)&=&\frac{1}{3}An_0\left[\frac{3}{2}-(\frac{1}{2}+x_0)I^2\right]u^2
+\frac{\frac{2}{3}Bn_0\left[\frac{3}{2}-(\frac{1}{2}+x_3)I^2\right]u^{\sigma+1}}
{1+\frac{2}{3}B'\left[\frac{3}{2}-(\frac{1}{2}+x_3)I^2\right]u^{\sigma-1}}
\nonumber \\ &+& u \sum_{i=1,2}\left[C_i \left({\cal J}^i_n+{\cal
J}^i_p\right)+\frac{(C_i-8Z_i)}{5}I\left({\cal J}^i_n-{\cal
J}^i_p\right)\right], \label{V-all}
\end{eqnarray}
where
\begin{equation}
{\cal J}^i_{\tau}(n,I,T)= \ 2 \int \frac{d^3k}{(2\pi)^3}
g(k,\Lambda_i)f_{\tau}. \label{J-tau}
\end{equation}

In Eq.~(\ref{V-all}), $I=(n_n-n_p)/n$ and $u=n/n_0$, with $n_0$
denoting the equilibrium symmetric nuclear matter density
$n_0=0.16$ fm$^{-3}$. The parameters $A$, $B$, $\sigma$, $C_1$,
$C_2$ and $B'$, which appear in the description of symmetric
nuclear matter and the additional parameters $x_0$, $x_3$, $Z_1$,
and $Z_2$ used to determine the properties of asymmetric nuclear
matter, are treated as parameters constrained by empirical
knowledge \cite{Prakash-97,Bombaci-01}.

The first two terms of the right-hand side of Eq.~(\ref{V-all})
arise from local contact nuclear interaction which led to power
density contributions such as in the standard Skyrme equation of
state. These are assumed to be temperature independent. The third
term describes the effects of finite range interactions, according
to the choice of the function $g(k,\Lambda_i)$, and is the
temperature dependent part of the interaction \cite{Bombaci-01}.
The function, $g(k,\Lambda_i)$, may have the following forms

\begin{enumerate}
\item Case 1:
\begin{equation}
 g_1(k,\Lambda_i)=\left[1+\left(\frac{k}{\Lambda_{i}}\right)^2
\right]^{-1}. \label{g-1} \end{equation}
In this case we introduce two finite range terms: one
corresponding to a long-range attraction and the other to a
short-range repulsion. The finite range parameters are
$\Lambda_1=1.5 k_F^{0}$ and $\Lambda_2=3 k_F^{0}$ and $k_F^0$ is
the Fermi momentum at the saturation point $n_0$. The function
$g_1(k,\Lambda_i)$ has been used extensively in previous papers
(see Refs.~\cite{Prakash-97,Bombaci-01} and references therein).

\item Case 2:

\begin{equation}
g_2(k,\Lambda_i)=\left[1-\left(\frac{k}{\Lambda_{i}}\right)^2
\right]. \label{g-2}
\end{equation}
In this case the finite range interactions are approximated by
effective local interactions by retaining only the quadratic
momentum dependence. Therefore, the energy density in
Eq.~(\ref{E-D-1}) takes the form of Skyrme's effective
interactions. Actually, as we will show later, the two functions
coincide, for low value of momenta $k$ ($k<1$ fm$^{-1}$), but they
exhibit different trends for high values of $k$.

\end{enumerate}

The energy density of asymmetric nuclear matter at density $n$ and
temperature $T$, in  good approximation, is expressed as
\begin{equation}
\epsilon(n,T,I)=\epsilon(n,T,I=0)+\epsilon_{sym}(n,T,I),
\label{e-asm-1}
\end{equation}
where
\begin{equation}
\epsilon_{sym}(n,T,I)=nI^2 E_{sym}^{tot}(n,T)=n I^2
\left(E_{sym}^{kin}(n,T)+E_{sym}^{int}(n,T)\right).
\label{e-sym-1}
\end{equation}
In Eq.~(\ref{e-sym-1}) the nuclear symmetry energy
$E_{sym}^{tot}(n,T)$, is separated in   two parts i.e.
$E_{sym}^{kin}(n,T)$ (kinetic) and $E_{sym}^{int}(n,T)$
(interaction).

From Eqs.~(\ref{e-asm-1}) and (\ref{e-sym-1}) and setting $I=1$,
we obtain that the nuclear symmetry energy $E_{sym}^{tot}(n,T)$ is
given by
\begin{equation}
E_{sym}^{tot}(n,T)=\frac{1}{n}\left(\epsilon(n,T,I=1)-\epsilon(n,T,I=0)
\right). \label{Esym-d-1}
\end{equation}
Thus, from Eqs.~(\ref{Esym-d-1}) and (\ref{E-D-1}) and a suitable
choice of the parameters $x_0$, $x_3$, $Z_1$ and $Z_2$, we can
obtain different forms for the density dependence of the symmetry
energy $E_{sym}^{tot}(n,T)$. It is well known that the need to
explore different forms for $E_{sym}^{tot}(n,T)$ stems from the
uncertain behavior at high density \cite{Prakash-97}. The
high-density behavior of symmetry energy is the least known
property of dense matter \cite{Kutschera-94,Li-02,Fuchs-06}, with
different nuclear models giving contradictory predictions. Thus,
in relativistic mean field (RMF) models, symmetry energy strongly
increases with the density of nuclear matter
\cite{Glendenning-97}, while in many realistic potential models of
nuclear matter in the variational approach \cite{Wiringa-88}, the
symmetry energy saturates and then bends over at higher densities.

Recently, the density dependence of the symmetry energy in the
equation of state of isospin asymmetric nuclear matter has been
studied using isoscaling of the fragment yields and the
antisymmetrized molecular dynamic calculation \cite{Shetty-07}. It
was observed that the experimental data at low densities are
consistent with the form of symmetry energy, $E_{sym}(u)\approx
31.6u^{0.69}$, in close agreement with those predicted by the
results of variational many-body calculations. In
Ref.~\cite{Shetty-07} it was suggested also that the heavy ion
studies favor a dependence of the form $E_{sym}(u)\approx
31.6u^{\gamma}$, where $\gamma=0.6-1.05$. This constrains the form
of the density dependence of the symmetry energy at higher
densities, ruling out an extremely "stiff" and "soft" dependence
\cite{Shetty-07}.

Additionally, by using the isospin dependent
Boltzmann-Uehling-Uhlenbeck transport model calculations, Chen et
al.~\cite{Chen-05} also showed that a stiff density dependence of
the symmetry energy parameterized as $E_{sym}(u)\approx
31.6u^{1.05}$ clearly explains the isospin diffusion data
\cite{Tsang-04} from NSCL-MSU (National Superconducting Cyclotron
Laboratory at Michigan State University).

In the present work, since we are interested mainly in the study
of thermal effects on the nuclear symmetry energy, we choose a
specific form for it, enabling us to accurately reproduce  the
results of many other theoretical studies
\cite{Lee-98,Sammarruca-08}. In Ref.~\cite{Lee-98} the authors
carried out a systematic analysis of the nuclear symmetry energy
in the formalism of the relativistic Dirac-Brueckner-Hartree-Fock
approach, using the Bonn one-boson-exchange potential. In a very
recent work \cite{Sammarruca-08}, the authors applied a similar
method to that in Ref.~\cite{Lee-98} for the microscopic
predictions of the equation of state in asymmetric nuclear matter.
In that case $E_{sym}(u)$ was obtained by employing the simple
parametrization $E_{sym}(u)=C u^{\gamma}$ with $\gamma=0.7-1.0$
and $C\approx 32$ MeV.  The authors concluded that a value of
$\gamma$ close to $0.8$ gives a reasonable description of their
predictions, although the use of different functions in different
density regions may be best for an optimal fit
\cite{Sammarruca-08}. The results of
Refs.~\cite{Lee-98,Sammarruca-08}  are well reproduced by
parameterizing the nuclear symmetry energy according to the
formula

\begin{equation}
E_{sym}^{tot}(n,T=0)= \underbrace{13
u^{2/3}}_{Kinetic}+\underbrace{17
F(u)}_{Interaction}.\label{Esym-3}
\end{equation}
For the function $F(u)$, which parametrizes the interaction part
of the SE, we apply the following three different cases
\begin{equation}
F_1(u)=\sqrt{u},\qquad F_2(u)=u, \qquad F_3(u)=\frac{2u^2}{1+u}.
\label{Fu-form}
\end{equation}
The parameters $x_0$, $x_3$, $Z_1$ and $Z_2$ are chosen so that
Eq.~(\ref{Esym-d-1}), for $T=0$, reproduces the results of
Eq.~(\ref{Esym-3}) for the three different forms of the function
$F(u)$. In addition, the parameters $A$, $B$, $\sigma$, $C_1$,
$C_2$ and $B'$ are determined  in order that $E(n=n_0)-mc^2=-16$
{\rm MeV}, $n_0=0.16$ fm$^{-3}$, and the incompressibility to be
$K=120,180,240$ {\rm MeV} for each of the three cases.

\subsection{Single particle potentials}
The single particle energy $e_{\tau}$, obtained by  the functional
derivative of the energy density (Eq.~(\ref{E-D-1})) with respect
to the distribution function $f_{\tau}$, is written as
\begin{equation}
e_{\tau}(n,I,k,T)=\frac{\hbar^2k^2}{2m}+U_{\tau}(n,I,k,T).
\label{esp-1}
\end{equation}

The single particle energy $e_{\tau}$ consists of a kinetic part
and an interaction one $U_{\tau}(n,I,k,T)$, which depends
explicitly on density, momentum, isospin asymmetry and temperature
as expected from an interaction term.

The single particle potential $U_{\tau}(n,I,k,T)$ (protons or
neutrons), obtained from the functional derivative of the
interaction part of the energy density (Eq.~(\ref{V-all})) with
respect to the distribution function $f_{\tau}$,  has the general
form
\begin{equation}
U_{\tau}(n,I,k,T)=U_{\tau}^A(n,I)+U_{\tau}^B(n,I)+U_{\tau}^{MD}(n,I,k,T).
\label{U-total}
\end{equation}

The first two terms are momentum independent, while the third one
describes the momentum dependence of the single particle
potential. The three terms have the following forms
\begin{equation}
U_{\tau}^A(n,I)=Au \mp \frac{2}{3}A(\frac{1}{2}+x_0)uI,
\label{UnA-tot}
\end{equation}
\begin{equation}
U_n^B(n,I)=\frac{U_{\tau}^{B1}(n,I)U_{\tau}^{B2}(n,I)-U_{\tau}^{B3}(n,I)U_{\tau}^{B4}(n,I)}
{\left[U_{\tau}^{B2}(n,I)\right]^2},\label{UnB-tot}
\end{equation}
with
\begin{eqnarray}
U_{\tau}^{B1}(n,I)&=&B(\sigma+1)u^{\sigma}\mp
\frac{4}{3}B(\frac{1}{2}+x_3)u^{\sigma}I+
\frac{2}{3}B(1-\sigma)(\frac{1}{2}+x_3)u^{\sigma}I^2, \nonumber
\\
U_{\tau}^{B2}(n,I)&=&1+\frac{2}{3}B'\left[\frac{3}{2}-(\frac{1}{2}+x_3)I^2\right]u^{\sigma-1},\nonumber
\\
U_{\tau}^{B3}(n,I)&=&\frac{B'}{n_0}(\sigma-1)u^{\sigma-2}\mp
\frac{4}{3}\frac{B'}{n_0}(\frac{1}{2}+x_3)u^{\sigma-2}I+
\frac{2}{3}\frac{B'}{n_0}(3-\sigma)(\frac{1}{2}+x_3)u^{\sigma-2}I^2,
\nonumber \\
U_{\tau}^{B4}(n,I)&=&\frac{2}{3}Bn_0\left[\frac{3}{2}-(\frac{1}{2}+x_3)I^2\right]u^{\sigma+1}
 \label{UB-all}
\end{eqnarray}
and
\begin{eqnarray}
U_{\tau}^{MD}(n,I,k,T)&=&\frac{4}{5}\frac{1}{n_0}
\sum_{i=1,2}\left[\frac{1}{2}\left(3C_i-4Z_i\right){\cal
J}^i_{\tau}+\left(C_i+2Z_i\right){\cal J}^i_{\tau'}\right]
\nonumber
\\
&+& u\sum_{i=1,2}\left[
\left(C_i\pm\frac{C_i-8Z_i}{5}I\right)g(k,\Lambda_i) \right]
\label{U-MD}.
\end{eqnarray}
The subscripts in the integrals are $\tau\neq \tau '$; the upper
signs stand for neutrons, while the lower ones for protons. An
advantage of the present model is that for $T=0$, the term
$U_{\tau}^{MD}(n,I,k,T)$, as well as all the quantities can be
derived in analytical forms.

It is of interest to see that the single particle potentials are
separated in two parts. The first one,
$U_{\tau}^A(n,I)+U_{\tau}^B(n,I)$ is a function only of the baryon
density $n$ and the isospin asymmetry parameter $I$. The second
one, $U_{\tau}^{MD}(n,I,k,T)$ has an additional dependence on $T$
and $k$. Actually, $U_{\tau}^{MD}(n,I,k,T)$ is mainly responsible
for the trend of the effective mass and also the effective mass
splitting. Additionally, it is connected with the effect of the
temperature on the interacting  part of the energy density. It is
also worthwhile to notice the direct correlation between the
regulator function $g(k,\Lambda_i)$ and $U_{\tau}^{MD}(n,I,k,T)$.
Thus, the choice of the function $g(k,\Lambda_i)$ plays an
important role for the single particle properties of hot nuclear
matter, but one expects it to also be  significant for the bulk
properties of nuclear matter.

\subsection{Nuclear symmetry potential}

The nuclear symmetry potential (NSP)  refers to the isovector part
of the nucleon mean-field potential in isospin asymmetric nuclear
matter, which in hot nuclear matter can also depend on the
temperature. Most of the studies concerning the NSP have been
carried out  for zero temperature, while the temperature
dependence of the NSP so far has received little theoretical
attention \cite{Xu-07-2}. The NSP potential, describes the
difference between the neutron and proton single particle
potentials in neutron rich matter and has the form
\begin{equation}
U_{sym}(n,I,k,T)=\frac{U_{n}(n,I,k,T)-U_p(n,I,k,T)}{2I}.
\label{Usym-1}
\end{equation}

Various theoretical models have been  applied to study the
symmetry potential. Most of them predict a symmetry potential
decreasing with increasing nucleon momentum. However, some nuclear
models which employed effective interaction, predict an opposite
behavior \cite{Xu-07-2}.

A systematic analysis of a large number of nucleon-nucleus
scattering experiments and (p,n) charge-exchange reactions at beam
energies up to $100$ MeV has shown that the data can be very well
described by the parametrization
\begin{equation}
U_{sym}(E_{kin})=a-bE_{kin}, \label{Usym-Lane}
\end{equation}
with $a \approx 22-34$ MeV and $b=0.1-0.2$. Actually, the
uncertainties for both parameters $a$ and $b$ are large. As
pointed out in Ref. \cite{Dalen-05}, the old analysis of Lane
\cite{Lane-62} is consistent with a decreasing trend of the
potential as function of $k$, while a more recent analysis based
on Dirac phenomenology \cite{Kozack-90} leads to the opposite
conclusions.

In order to clarify the effects of the momentum dependence on the
NSP, it is easy to show, by applying
Eqs.~(\ref{U-total})-(\ref{U-MD}), that $U_{sym}(n,I,k,T)$, at
zero temperature is
\begin{equation}
U_{sym}(n,I,k)=U_{sym}^{MIP}(n,I)+U_{sym}^{MDP}(n,k).
\label{Usym-2}
\end{equation}
In Eq.~(\ref{Usym-2}), $U_{sym}^{MIP}(n,I)$ is  momentum
independent, while  the momentum dependent part
$U_{sym}^{MDP}(n,k)$ is written as
\begin{equation}
U_{sym}^{MDP}(n,k)=\frac{u}{5} \sum_{i=1,2}\left(C_i-8Z_i\right)
g(k,\Lambda_i) . \label{Usym-MDP}
\end{equation}
From Eq.~(\ref{Usym-MDP}) it is obvious that the behavior of
$U_{sym}(n,k)$ as a function of k is defined from the values of
the parameters $C_i$ and $Z_i$, as well as from the function
$g(k,\Lambda_i)$. In fact, the regulator function $g(k,\Lambda_i)$
specifies the trend of the slope. Thus, it is interesting to study
the effect of the parametrization, including the function $F(u)$,
and the contribution of the choice of a specific function
$g(k,\Lambda_i)$ on the properties of the NSP, both for cold and
hot nuclear matter.

\subsection{Effective mass}
The nucleon effective mass is one of the most important
single-particle properties of nuclear matter. It characterizes the
momentum dependence of the single-particle potential of a nucleon,
and  consequently  the quasiparticle properties of a nucleon
inside a strongly interacting medium such as the nuclear matter.
Moreover, the effective mass describes to leading order the
effects related to the nonlocality of the underlying nuclear
effective interaction and the  Pauli exchange effects in
many-fermion systems \cite{Dalen-05,BLi-04,Lesinski-06}.

The effective mass $m_{\tau}^{*}(k)$, is determined by the
momentum-dependent single nucleon potential via
\begin{equation}
\frac{m_{\tau}^{*}(n,I,k)}{m_{\tau}}=\left[1+\frac{m_{\tau}}{\hbar^2
 k}\frac{dU_{\tau}(n,I,k,T)}{dk}\right]^{-1}. \label{effe-1}
\end{equation}
According to Eq.~(\ref{effe-1}), $m_{\tau}^{*}(n,I,k)$ generally
depends on the baryon density, the isospin asymmetry and the
momentum of the nucleon. In the present model $m_{\tau}^{*}$ is
independent of $T$.

An evaluation of $m_{\tau}^{*}$ mass at the Fermi momentum
$k=k_F$, employing Eq.~(\ref{effe-1}), yields the Landau effective
mass.
An advantage of the present model is  that by applying
Eq.~(\ref{U-total}), we get $m_{\tau}^{*}$ in  analytical form
i.e.
\begin{equation}
\frac{m_{\tau}^{*}(n,I)}{m_{\tau}}=\left[1+u\frac{m_{\tau}}{\hbar^2
 k_{F}}\sum_{i=1,2}
\left(C_i\pm\frac{C_i-8Z_i}{5}I\right)\frac{dg(k,\Lambda_i)}{dk}\mid_{k=k_F}
\right]^{-1}. \label{Ef-mas-3}
\end{equation}
Eq.~(\ref{Ef-mas-3}) exhibits the dependence of the effective mass
on the values of the parameters $C_i$, $Z_i$ and $\Lambda_i$ as
well as on the derivative of the regulator function
$g(k,\Lambda_i)$. In particular, for the two cases of the function
$g(k,\Lambda_i)$ (given in Eqs.(\ref{g-1}), and (\ref{g-2})), we
get respectively for the effective masses
\begin{equation}
\frac{m_{\tau}^{*}(n,I)}{m_{\tau}}=\left[1-\frac{2 u
m_{\tau}}{\hbar^2}\sum_{i=1,2}\frac{1}{\Lambda_i^2}\frac{
\left(C_i\pm\frac{C_i-8Z_i}{5}I\right)}{\left[1+\left(\frac{k_F^0}{\Lambda_i}
\right)^2\left((1\pm I)u\right)^{2/3} \right]^2} \right]^{-1},
\qquad case-1 \label{mef-g1}
\end{equation}
\begin{equation}
\frac{m_{\tau}^{*}(n,I)}{m_{\tau}}=\left[1-\frac{2 u
m_{\tau}}{\hbar^2}\sum_{i=1,2}\frac{1}{\Lambda_i^2}
\left(C_i\pm\frac{C_i-8Z_i}{5}I\right)\right]^{-1}. \qquad case-2
\label{mef-g2}
\end{equation}

The meaning of Eqs.~(\ref{mef-g1}) and (\ref{mef-g2}) is clear.
Firstly, the Landau effective mass for protons and neutrons
depends on the parameters $C_i$ and $Z_i$, where $C_i$ is
connected with the saturation properties of nuclear matter and
$Z_i$ with the density dependence of the nuclear symmetry energy.
Secondly, $m_{\tau}^{*}$ depends on the baryon density via the
variable $u$, as well as on the isospin parameter $I$. Finally,
there is also a direct dependence of  $m_{\tau}^{*}$ on the
regulator function $g(k,\Lambda_i)$ via the parameters
$\Lambda_i$. Hence, one can conclude that  the values of the
effective mass are reflected on the specific properties of the
applied nuclear models. This enables us to carry out  a systematic
study of the properties of $m_{\tau}^{*}$ by applying the
parametrization of the present model.


\section{Results and Discussion}

We have studied thermal and momentum-dependent effects on the
properties of hot asymmetric nuclear matter including mean field
potentials (proton, neutron and symmetry) as well as the effective
mass, by applying a momentum-dependent effective interaction. The
parametrization of the model can be found in Table~1 and Table~2
of Ref.~\cite{Prakash-97}.


In particular, we have studied two general cases. In the first one
(called $g_1$) the momentum regulator function $ g(k,\Lambda_i)$
is given by equation (\ref{g-1}). In this case, the quantity
${\cal J}_{\tau}^i$ defined by Eq.~(\ref{J-tau}), at $T=0$, takes
the form
\begin{equation}
{\cal
J}_{\tau}^{i(g_1)}(n,I)=\frac{3}{2}n_0\left(\frac{\Lambda_i}{k_F^0}\right)^3
\left(\frac{\left((1\pm
I)u\right)^{1/3}}{\frac{\Lambda_i}{k_F^0}}- \tan^{-1}
\frac{\left((1\pm
I)u\right)^{1/3}}{\frac{\Lambda_i}{k_F^0}}\right),
\label{Inp-Case-1}
\end{equation}
where the upper signs refer to neutrons and the lower ones to
protons.

In the second case (called $g_2$) the momentum regulator function
$ g(k,\Lambda_i)$ is given by equation (\ref{g-2}). Here, the
finite range interactions are approximated as effective local
interactions by retaining only the quadratic momentum dependence
\cite{Prakash-97,Lee-01}. In this case, the energy density takes
the form of Skyrme's effective interaction. At $T=0$ the quantity
${\cal J}_{\tau}^i$ is
\begin{equation}
{\cal J}_{\tau}^{i(g_2)}(n,I)=\frac{9}{2}n_0\left((1\pm I) u
\right)^{1/3} \left(1- \frac{3}{5}\frac{(\left(1\pm
I)u\right)^{2/3}}{(\frac{\Lambda_i}{k_F^0})^2} \right).
\label{In-Case-2-2}
\end{equation}

The parameters $A$, $B$, $\sigma$, $C_1$, $C_2$ and $B'$ (a small
parameter introduced  to maintain causality), are determined from
constraints provided by the empirical properties of symmetric
nuclear matter at the equilibrium density $n_0=0.16$ fm$^{-3}$.
With the appropriate choice of the parameters, it is possible to
parametrically vary the nuclear incompressibility $K$, so that the
dependence on the stiffness of the equation of state may be
explored. In the same way, by  choosing the appropriate parameters
$x_0$, $x_3$, $Z_1$ and  $Z_2$, it is  possible to obtain
different forms for the density dependence of the nuclear symmetry
energy~\cite{Prakash-97,Bombaci-01}.

Fig.~1(a) displays the behavior of the NSE as a function of the
ratio $u=n/n_0$ for the three different parametrizations of the
function $F(u)$ (relation [\ref{Fu-form}]). The function $F_3(u)$
leads to a stiffer nuclear symmetry energy dependence on the
density, while the function $F_1(u)$  leads to a softer one. It is
worthwhile to point out that the above parametrization of the
interacting part of the nuclear symmetry energy is extensively
used for the study of neutron star properties
\cite{Prakash-97,Bombaci-01}, as well as for the study of the
collisions of neutron-rich heavy ions with intermediate energies
\cite{Baran-05,Li-97}. For a very recent review of the
applications of the proposed momentum dependent effective
interaction model and the specific parametrization of it, see
Ref.~\cite{Li-08} (and references therein). The aim of the above
simple parametrization is to reproduce the nuclear symmetry energy
originating from various theoretical calculations and/or
experimental predictions and also to be able to cover the possible
range of the nuclear symmetry energy dependence on the density.

In Fig.~1(b) we plot the regulator momentum function
$g(k,\Lambda_1)$ for the two cases applied in the present work.
The most striking feature of the two curves is the fact that for
low values of the momenta $k$, the two cases coincide, but for
higher values of $k$ they have completely different behavior. The
motivation for applying these two different parameterizations for
the momentum part of the interaction is to investigate in greater
detail the momentum dependence effects on the mean field
properties of hot asymmetric matter.

The nuclear symmetry potential, pertaining to the isovector part
of the nucleon mean field potential of a nucleon in nuclear
matter, also depends on the momentum of the nucleon. In order to
see the temperature and momentum effects on the nuclear symmetry
potential, we first study  the above effects on the nucleon
single-particle potentials in hot  asymmetric nuclear matter.

In Fig.~2 we present  the behavior of the proton and neutron
single particle potentials for the case $g_1$ and three different
values of the incompressibility $K$. The calculations are
performed at the saturation density $n_0$. In each figure we
display $U_p$ and $U_n$ for three different forms of the symmetry
energy (named 1, 2 and 3 respectively, see Eq.~[\ref{Fu-form}]).
It is noted that in case 3, for $K=180$ MeV and for high values of
$k$, $U_p$ and $U_n$ obtain comparable values. As we will see
later, this affects the behavior of NSP. A feature of Fig.~2 is
the effect of $K$ on the values of the SPP for high values of $k$.
Thus, for small values of $k$ in the three cases, both for protons
and neutrons, lead to similar values for the SPP, while for high
values of $k$, there is an obvious splitting of the values of SPP,
which shows a monotonic increase with $K$.

Fig.~3 displays the trend of $U_p$ and $U_n$, in case $g_1$, for
$K=240$ and $F_2(u)=u$ as a function of the asymmetry parameter
$I=(n_n-n_p)/n$. The meaning of the trend is very clear. The
increase of the isospin asymmetry parameter $I$ leads to a more
attractive $U_p$ and also more repulsive $U_n$.  Similar behavior
is found in the other cases as well.

In Fig.~4 we plot $U_{p,n}$ for the case $g_2$ and for the three
different values of $K$. The competitive behavior of $U_{p}$ and
$U_n$, both for low and high values, is clearly reflected  in the
behavior of $U_{sym}$ as indicated later in Fig.~7.

The temperature effects on $U_p$, $U_n$ are shown in Fig.~5. As
expected, the increase of $T$ leads to a corresponding increase of
the values of $U_p$ and $U_n$. The effects are more pronounced for
$T>5$ MeV. The temperature effect on $U_p$, $U_n$ has a
significant influence on the temperature dependence of the nuclear
symmetry energy as found in  previous works
\cite{Moustakidis-07,Xu-07-2}.

Most of the theoretical models predict a decreasing symmetry
potential with increasing nucleon momentum, albeit at different
rates, while a few  nuclear effective interactions used in some
models show  opposite behavior. In the present work we try to
clarify the above controversial results by applying a systematic
study of $U_{sym}$.  In Fig.~6, we plot the nuclear symmetry
potential $U_{sym}=(U_n-U_p)/2I$ as a function of the nucleon
kinetic energy $E_{kin}$ in case $g_1$, for three values of $K$
and in each case for three different choices of $F(u)$. In the
present work, we apply $I=0.4$ in the calculation of $U_{sym}$ and
as  pointed out also in \cite{Xu-07-2,Dalen-05}, $U_{sym}$ is
almost independent of the asymmetry parameter $I$ but depends
strongly on density and momentum. It is of interest to see that
for the nine, in total, different cases, only in two of them
$U_{sym}$ decreases with increasing nucleon momentum ( for $K=180$
MeV and $F_3(u)=2u^2/(u+1)$ and for $K=240$ MeV and
$F(u)=u^{1/2}$). This trend is in agreement with experiment, which
is well reproduced using the empirical relation (\ref{Usym-Lane}).
In the remaining seven cases, $U_{sym}$ increases with increasing
nucleon momentum. In addition, from Fig.~6 it is concluded that
the energy dependence of $U_{sym}$ is sensitive to the
incompressibility $K$.

 It is worthwhile to notice that the nuclear symmetry
potentials differs from the nuclear symmetry energy as the latter
involves the integration of the isospin-dependent mean-field
potential of a nucleon over its momentum \cite{Xu-07-2}. However,
it is of interest to study the effect of the potential part of
$E_{sym}(u)$ on the momentum and density dependence of $U_{sym}$.
Fig.~6 demonstrates the strong dependence of $U_{sym}$ on the
function $F(u)$. It seems that by applying the functions
$F(u)=u^{1/2}$ and $F(u)=u$ we receive similar results only for
low values of $E_{kin}$ (except for $K=120$ MeV where  a similar
trend is taken also for higher values of $E_{kin}$). The case with
$F(u)=2u^2/(u+1)$, which introduces a much stiffer density
dependence $E_{sym}$, leads to different results even for low
values of momenta. To sum up, the density dependence of $E_{sym}$
is  well reflected on the momentum dependence of $U_{sym}$, so it
is expected that models with different density dependence in the
nuclear symmetry energy may predict different energy dependence of
$U_{sym}$.

In Fig.~7 we also plot  $U_{sym}$ as a function of $E_{kin}$, for
the case $g_2$. In fact, the trends for $K=120$ MeV and $K=240$
MeV coincide, and $U_{sym}$ exhibits a negative slope, while  for
$K=180$ MeV the slope of $U_{sym}$ is positive.

The key quantity to explain the trend of $U_{sym}$ as a function
of $E_{kin}$ is expression (\ref{Usym-MDP}). For this expression
we conclude that the behavior of $U_{sym}(n,k)$ as a function of
$k$ is defined from the values of the parameters $C_i$ and $Z_i$
as well as from the function $g(k,\Lambda_i)$. Both the parameters
$C_i$ and $Z_i$ are related to the strength of the momentum
dependence. In addition, $C_i$ is fixed also to reproduce the
properties of  symmetric nuclear matter at the saturation point,
while $Z_i$ is set to fix the density dependence of the symmetry
energy. The regulator function $g(k,\Lambda_i)$ mainly affects the
trend of the slope of   $U_{sym}$. In particular, in case $g_2$,
after some algebra, the expression (\ref{Usym-MDP}) can be written
as
\begin{equation}
U_{sym}^{MDP}(n,k)=U_{sym}^{MI}(n,C_i,Z_i)-{\cal D}_1  E_{kin} u
\sum_{i=1,2}\frac{\left(C_i-8Z_i\right)}{\Lambda_i^2} ,
\label{Usym-MDP-1}
\end{equation}
where ${\cal D}_1$ is a constant.  Eq.~(\ref{Usym-MDP-1})
establishes a linear relation between $U_{sym}$ and $E_{kin}$, in
case $g_2$, as  indicated in Fig.~9.

In case $g_1$ for low values of $k$, a similar relation as
(\ref{Usym-MDP-1}) holds. Accordingly, for high values of $k$ we
find
\begin{equation}
U_{sym}^{MDP}(n,k)\simeq \frac{{\cal D}_2} {E_{kin}} u
\sum_{i=1,2}\left(C_i-8Z_i\right)\Lambda_i^2 , \label{Usym-MDP-2}
\end{equation}
where ${\cal D}_2$ is a constant.

From the above analysis it is clear that the behavior of $U_{sym}$
depends on the specific values of the parameters $C_i$, $Z_i$ and
$\Lambda_i$ and also on combinations of them. In particular, the
values of the parameters define the sign of the slope (positive or
negative) and the regulator function $g(k,\Lambda_i)$ defines the
trend of the slope.  Thus, the role of the regulator function
$g(k,\Lambda_i)$ is of importance, concerning the
momentum-dependent behavior of the proton and neutron single
particle potential, as well as the symmetry potential.

In view of the above comment it may be interesting to explore
other choices for the regulator function as those used in Ref.
\cite{Bogner-07}.
Such  work is in progress.

It is worth noting that according to Fig.~7, $U_{sym}$ is strongly
related
 to the density dependence  of the nuclear symmetry
energy and the total parametrization of the interaction energy
density part, i.e. for various values of the incompressibility
$K$.

The three cases where $U_{sym}$ is a decreasing function of
nucleon momentum are shown in Fig.~8 and compared with the
phenomenological expression (\ref{Usym-Lane}).

In Fig.~9 we plot $U_{sym}$ as a function of $E_{kin}$ for the
nuclear symmetry energy parametrization $F_1(u)=u^{1/2}$ and
$F_2(u)=u$ and for three different values of $u$. In both cases,
$U_{sym}$, exhibits  strong density dependent behavior. In
particular, there is a monotonic increase of $U_{sym}$ as function
of the nucleon density.

Special effort has been devoted to the study of the thermal
effects on $U_{sym}$. So far, that problem has received  little
theoretical attention  \cite{Xu-07-2}. Temperature influences  the
first term of the right-hand side of Eq.~(\ref{Usym-2}) i.e.
$U_{sym}^{MIP}(n,I)$.  The thermal effects on $U_{sym}$ are
displayed in Fig.~10. In the first case (Fig.~10[a]) an increase
of $T$ leads to decreasing values of $U_{sym}$, while the inverse
behavior is seen in the second case (Fig.10[b]). However, in both
cases thermal effects do play a role, especially for $T>5$.
Temperature does not change the trend, but only slightly affects
the values of $U_{sym}$ depending on $E_{kin}$.

Another quantity of interest, which can be easily calculated, is
the isoscalar potential.  Several decades ago, it was already
pointed out that the quantity $(U_n+U_p)/2$, which is obviously
the single-nucleon potential in absence of asymmetry, should be a
reasonable approximation to the isoscalar part of the optical
potential. The momentum dependence of $(U_n+U_p)/2$ is important
for extracting information about the symmetric matter equation of
state \cite{Sammarruca-05}. Now, in order to check the validity of
the model parameters this is customary, and a more stringent test
to compare the isoscalar potentials, as calculated in the present
work, with them of the variational many-body (VMB) predictions by
Wiringa \cite{Wiringa-88,BLi-04}.

In Fig.~11 we plot the isoscalar potential at the four values of
momenta $k$ ($k=1,2,3,4$ fm$^{-1}$) for four different cases. The
results of the present work compared with those (named UV14+UVII)
predicted by Wiringa~\cite{Wiringa-88}. In Fig.~11(a) indicated
that for the specific case, the predictions of the present work
are in  very good agreement with the VMB predictions up to about
$k=3$ fm$^{-1}$. In a second case, presented in Fig.~11(b), there
is agreement only for low values of the density and for momenta
$k$. There is  obvious disagreement for the predictions presented
in Figs.~11(c) and~11(d). From the above analysis it is concluded
that by suitable choice of the model parameters, we are able to
reproduce and/or  be  consistent with the predictions of other
microscopic many-body calculations or experimental constraints.
This is an advantage of the present model that is flexible enough
to reproduce predictions of  many other theoretical models.


The Landau effective mass splitting is shown in Fig.~12, where the
effective masses of proton and neutron are displayed as functions
of the asymmetry parameter $I$ for all cases corresponding to
$g_1$. For all cases,  $m^{*}_p$ is a decreasing function of $I$,
while $m^{*}_n$ is an increasing function of $I$. The only
exception is the case for $K=240$ MeV with $F_3(u)=2u^2/(u+1)$,
where $m^{*}_n$ is a decreasing function of $I$. In every case the
isospin mass splitting ($m^*_n-m^*_p$) is positive and is directly
dependent on the incompressibility $K$ and the parametrization
function $F(u)$.

In the present work, we also address   the problem which arises
when the effective mass splitting directly affects  the energy
dependence of the symmetry potential \cite{Baran-05}. We found
that in the framework of the proposed model and for the regulator
function $g_1$, all the cases lead to a positive splitting but the
slope of the $U_{sym}$ may be positive or negative. So, it is
concluded that the trend of $U_{sym}$, by applying the present
parametrization, does not directly connect to the effective mass
splitting. Nonetheless, we found that there is a correlation
between $U_{sym}$ and $m^*_n-m^*_p$ originated from the density
dependence of $E_{sym}$. More precisely, the cases with
$F(u)=u^{1/2}$ and $F(u)=u$ which lead to much closer values for
$U_{sym}$ also lead  to a similar value of $m^*_n-m^*_p$ as is
displayed in Figs.~12 and ~13.

To illustrate further the dependence of the effective mass on the
asymmetry parameter $I$ and to find the quantitative
characteristic of this dependence, the values of $m^{*}_{\tau}/m$
for various values of $I$ have been derived with the least-squares
fit method and found to take the general form
\begin{equation}
\frac{m^{*}_{\tau}(I)}{m} \simeq c_0+c_1I+c_2I^2. \label{m-I-fit}
\end{equation}

The parameters $c_i$ depend on the model parameters
$C_i,Z_i,\Lambda_i$ and they are different for each case.
Specifically, for low values of $I$ it is obvious that  linear
relations between $m^{*}_{\tau}/m$  and $I$ is hold (see also
Fig.~12). The splitting of effective masses, according to
Eq.~(\ref{m-I-fit}), is well approximated by
\begin{equation}
\Delta m^{*}(I)=m^{*}_{n}(I)-m^{*}_{p}(I)\simeq d_1I+d_2I^2,
\label{m-I-split}
\end{equation}
where  the parameters $d_i$ also depend on the model parameters
$C_i,Z_i,\Lambda_i$. For low values of $I$, the linear relation
$\Delta m^{*}(I)\approx d_1I$ is hold.

The effective mass as a function of $u$ is indicated in Fig.~13.
In all cases the effective mass splitting  ($m^{*}_n- m^{*}_p$) is
positive.

In Fig.~14(a) the effective mass, for the case $g_2$, is
displayed. The splitting  ($m^{*}_n- m^{*}_p$) is positive for the
case with $K=120$ MeV (and also for $K=240$ MeV) and negative for
$K=180$ MeV. Thus, one can conclude by also comparing with the
case $g_1$, that the regulator function $g(k,\Lambda_i)$ can have
a dramatic effect on the effective mass splitting.  The above
feature is displayed also in Fig.~14(b), where the effective mass
is plotted as a function of $u$. It is worth  pointing out that in
case $g_2$, an almost linear relation between $m^{*}_{\tau}/m$ and
$I$ is hold even for high values of $I$ (in comparison with the
case $g_1$). So, the relation (\ref{m-I-fit}) with $c_2=0$,
clearly describes the above dependence.

It should be emphasized that according to Figs.~12 and~13, the
density dependence of the symmetry energy influences the values
and also the slope of the effective mass (both for proton and
neutron), but does not affect the sign of the mass splitting,
which always remains positive (case $g_1)$. Nevertheless, in case
$g_2$, for different values of $K$ the sign of the mass splitting
may be negative or positive. In the case with $K=180$ the positive
slope of $U_{sym}$ (see Fig.~7) is connected with the negative
splitting (see Fig.14(a)). This result is in accordance with the
finding of Ref.~\cite{BLi-04,Baran-05}.

From the above analysis one can conclude that, in general, and by
applying the proposed, in the present  momentum effective
interaction model,  the slope (positive or negative ) of the
$U_{sym}$ as a function of $E_{kin}$ is not connected directly
with the sign of the effective mass splitting. Certainly, there is
a connection between the above quantities in the sense that both,
according to Eqs.~(\ref{mef-g1}), (\ref{mef-g2}) and
(\ref{Usym-MDP}), depend in a similar way on the parameters $C_i$,
$Z_i$ $\Lambda_i$.

\section{Summary}
In this study we have applied a momentum dependent effective
interaction in order to investigate the single-particle properties
of hot asymmetric nuclear matter. More specifically, we have
examined  the single-particle potentials of protons and neutrons,
the asymmetry potential and the isospin mass splitting for various
cases. The effects of the specific parametrization of the
interaction part of the energy on the single-particle properties
are studied and analyzed. It has been concluded that the behavior
of the symmetry potential depends strongly on the parametrization
of the interaction part of the energy density as well as on the
momentum dependence of the regulator function. The effects of the
parametrization are less pronounced on the isospin mass splitting.
The effect of an increase of the temperature is just to shift
higher the values of the proton and neutron single particle
potential. The symmetry potential $U_{sym}$ can be an increasing
or decreasing function of the nucleon kinetic energy, depending on
the parametrization of the momentum dependent effective
interaction model. In the first case $U_{sym}$ increases with the
temperature, while in the second one the inverse behavior is
observed.

\section*{Acknowledgments}
The author would like to thank Prof. S.E. Massen and Dr. C.P.
Panos for useful comments on the manuscript. The work was
supported by the Pythagoras II Research project (80861) of
E$\Pi$EAEK and the European Union.


\newpage

\begin{figure}
\centering
\includegraphics[height=8.0cm,width=8.0cm]{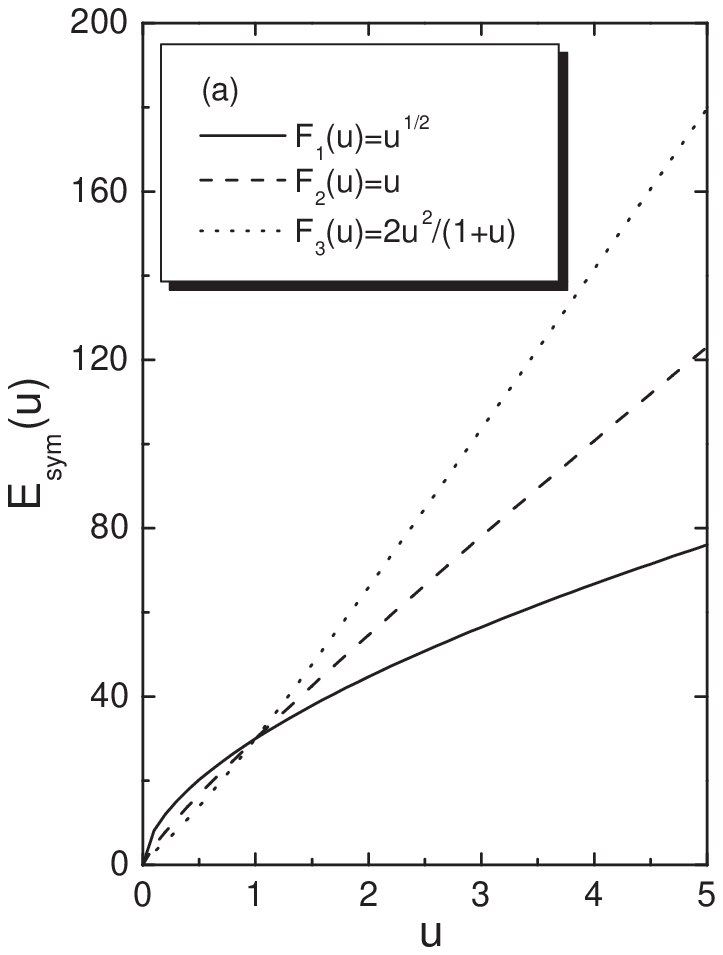}\
\includegraphics[height=8.0cm,width=8.0cm]{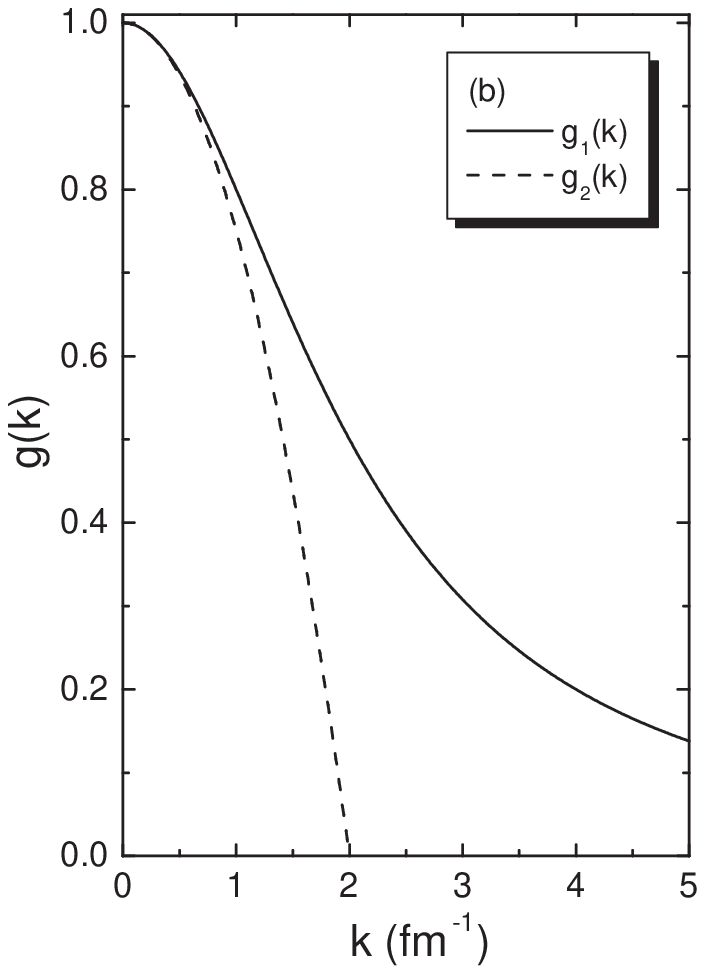}
\caption{(a) The nuclear symmetry energy as a function of
$u=n/n_0$ for three different parametrizations of the function
$F(u)$ given by Eq.~ (\ref{Fu-form}). (b) The regulator momentum
function
 $g(k,\Lambda)$ for the two cases, given by Eqs.~(\ref{g-1}) and (\ref{g-2}), applied in the present
 work. For more details see text.  } \label{}
\end{figure}
\begin{figure}
\centering
\includegraphics[height=5.5cm,width=5.5cm]{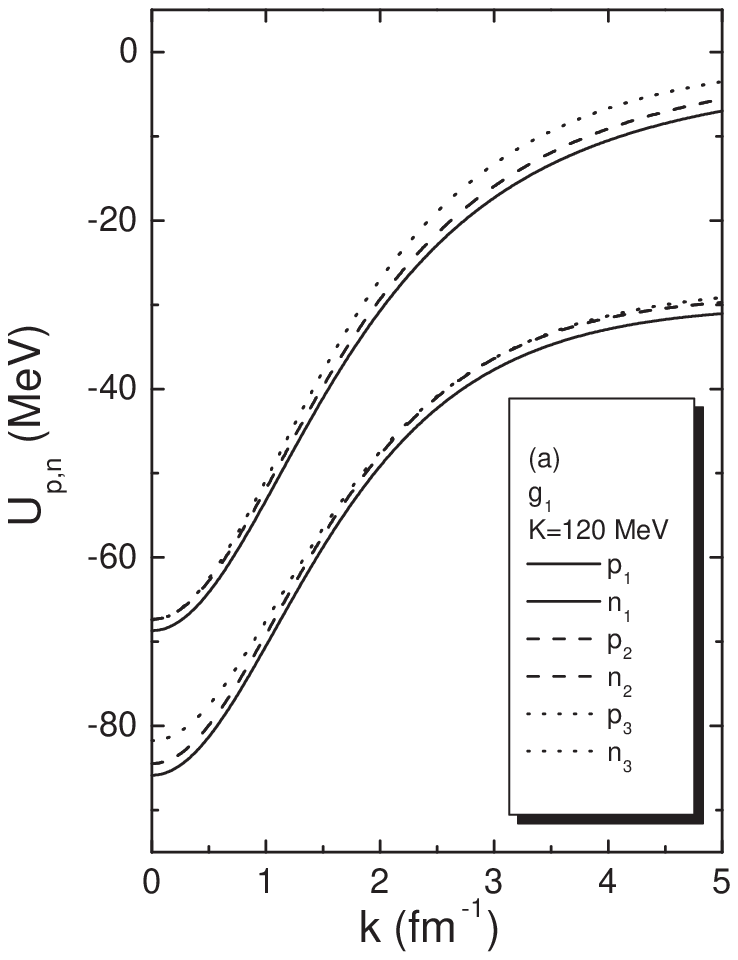}\
\includegraphics[height=5.5cm,width=5.5cm]{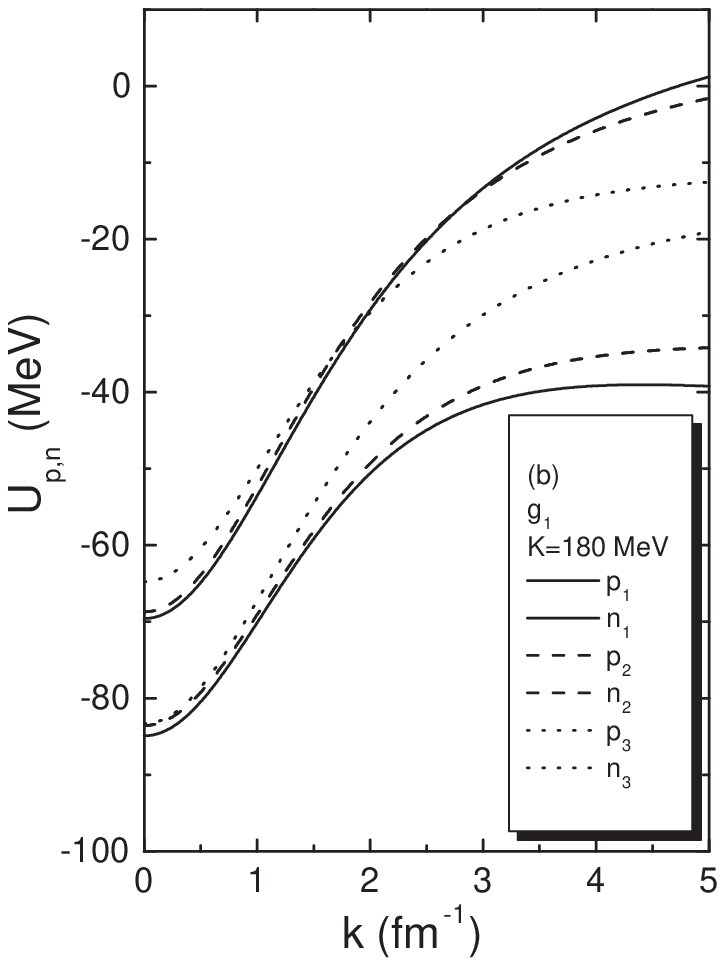}\
\includegraphics[height=5.5cm,width=5.5cm]{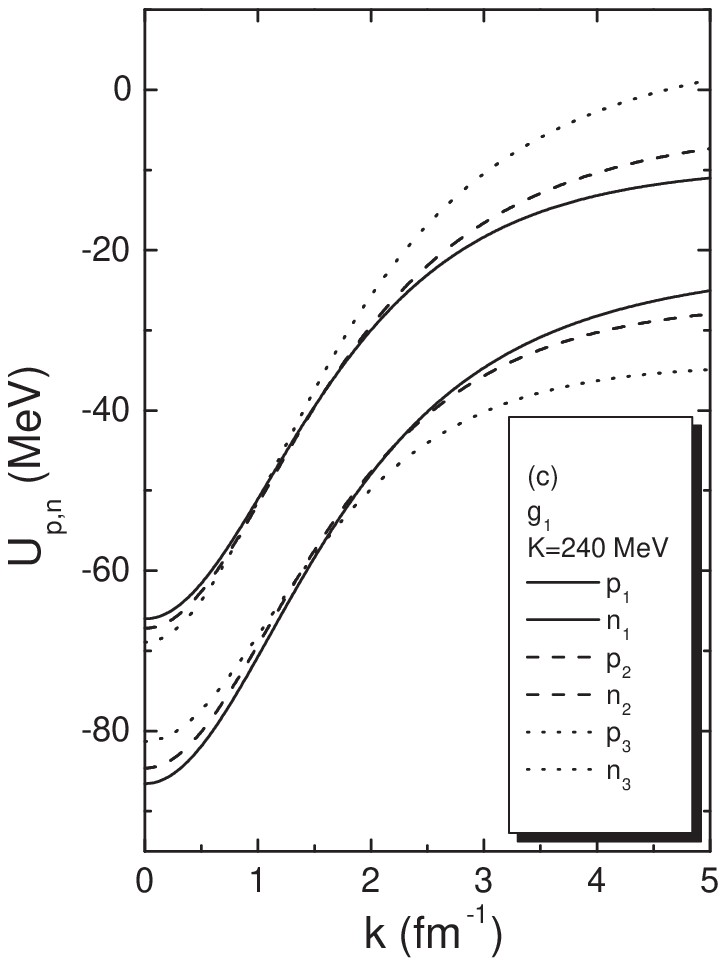}
\caption{ The proton and neutron single particle potentials (the
upper curves corresponds to $U_n$, while the lower one to $U_p$)
for the case $g_1$ and for three different vales of the
incompressibility $K$ (at the saturation density $n_0$). In each
figure we display $U_p$ and $U_n$ for three different forms of the
nuclear symmetry energy (named 1,2 and 3 respectively). } \label{}
\end{figure}

\begin{figure}
\centering
\includegraphics[height=8.0cm,width=8.0cm]{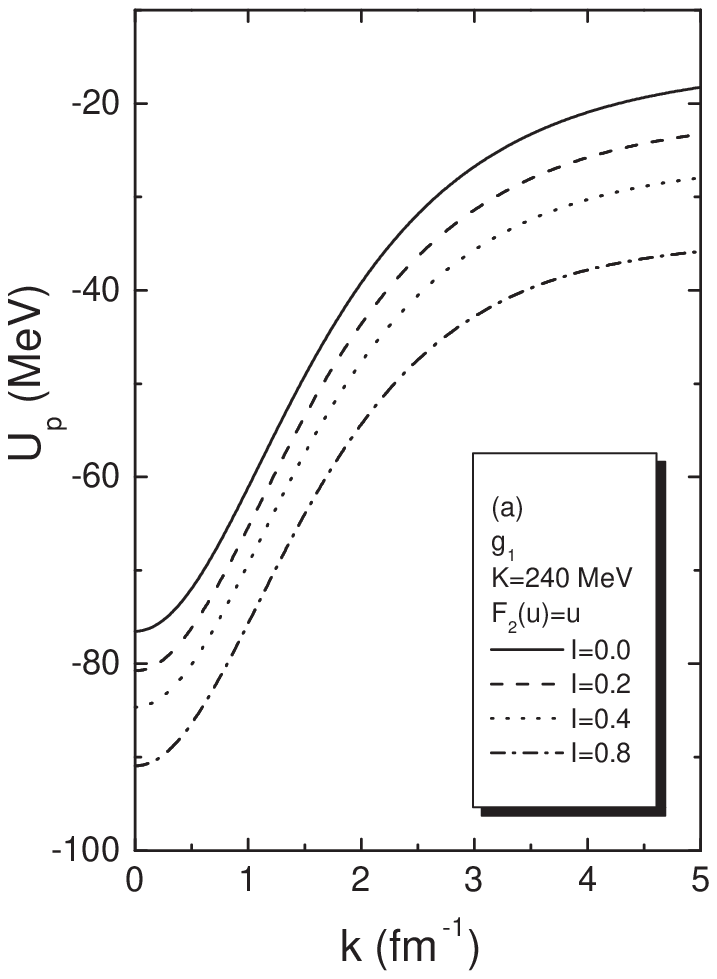}\
\includegraphics[height=8.0cm,width=8.0cm]{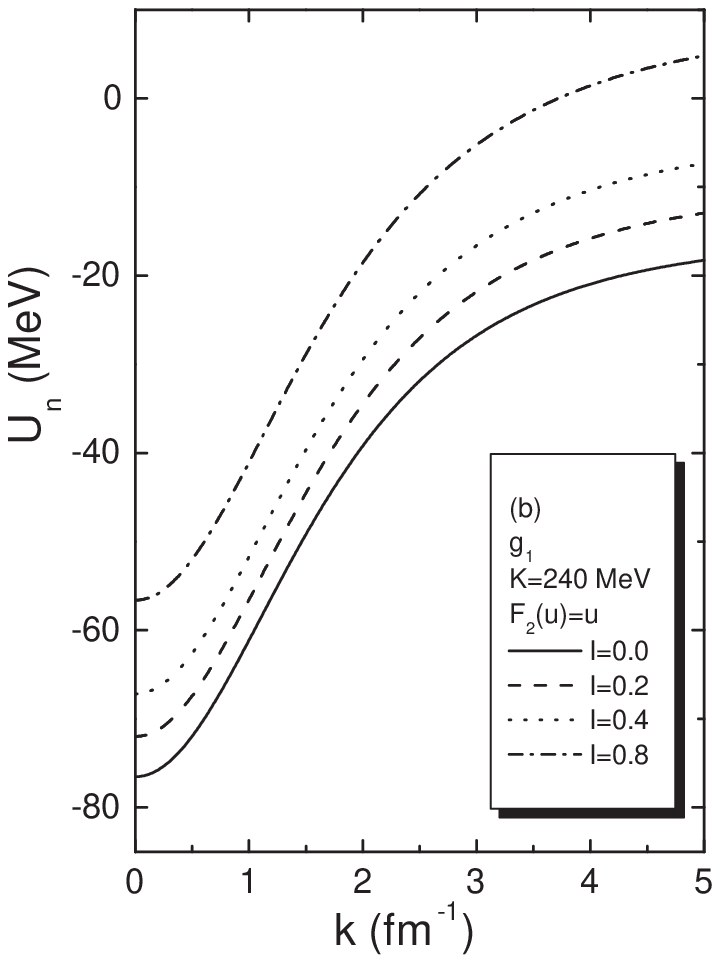}
\caption{$U_p$ and $U_n$, in case $g_1$ and for $K=240$ MeV and
$F_2(u)=u$ versus $k$ and for various values of the asymmetry
parameter $I=(n_n-n_p)/n$. } \label{}
\end{figure}
\begin{figure}
\centering
\includegraphics[height=8.0cm,width=8.0cm]{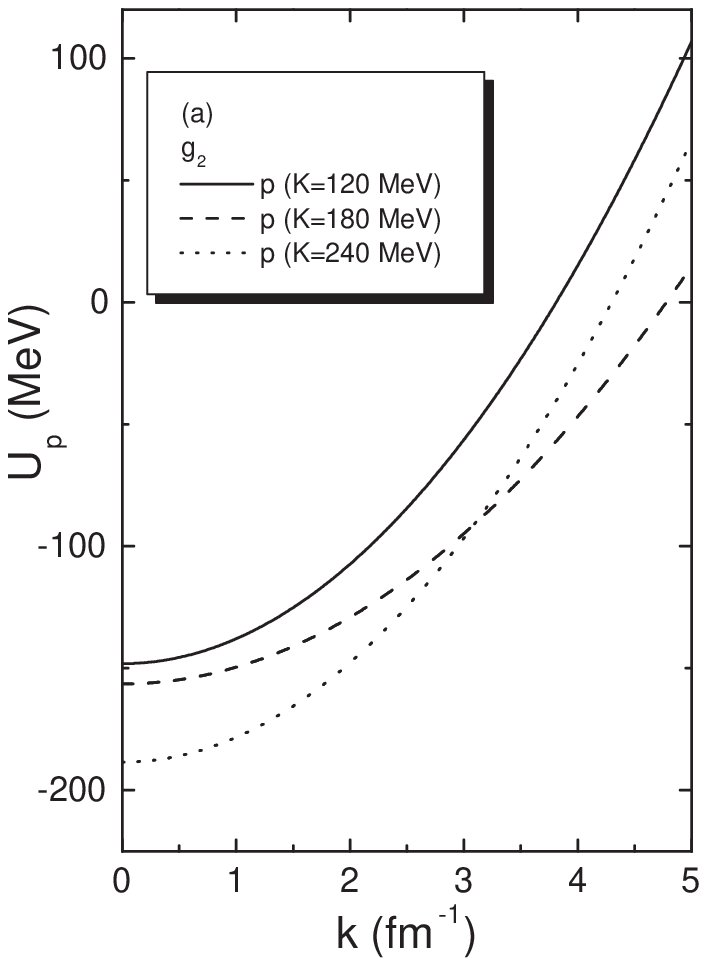}\
\includegraphics[height=8.0cm,width=8.0cm]{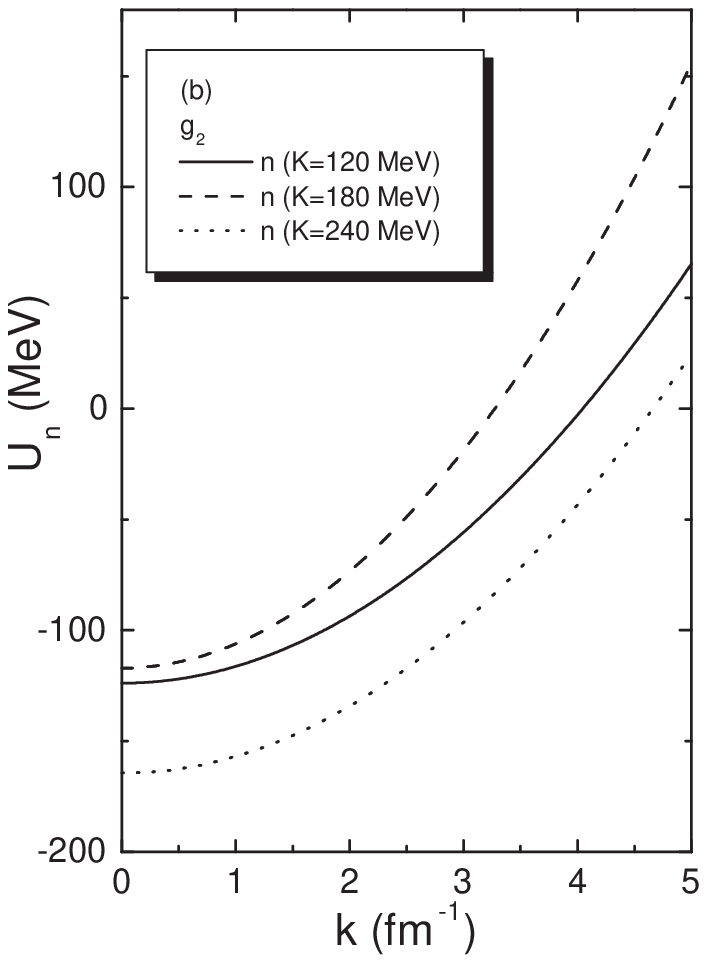}
\caption{$U_{p}$ (left) and $U_n$ (right) for the case $g_2$ and
for  three different values of $K$ (at the saturation density
$n_0$). } \label{}
\end{figure}
\begin{figure}
\centering
\includegraphics[height=8.0cm,width=8.0cm]{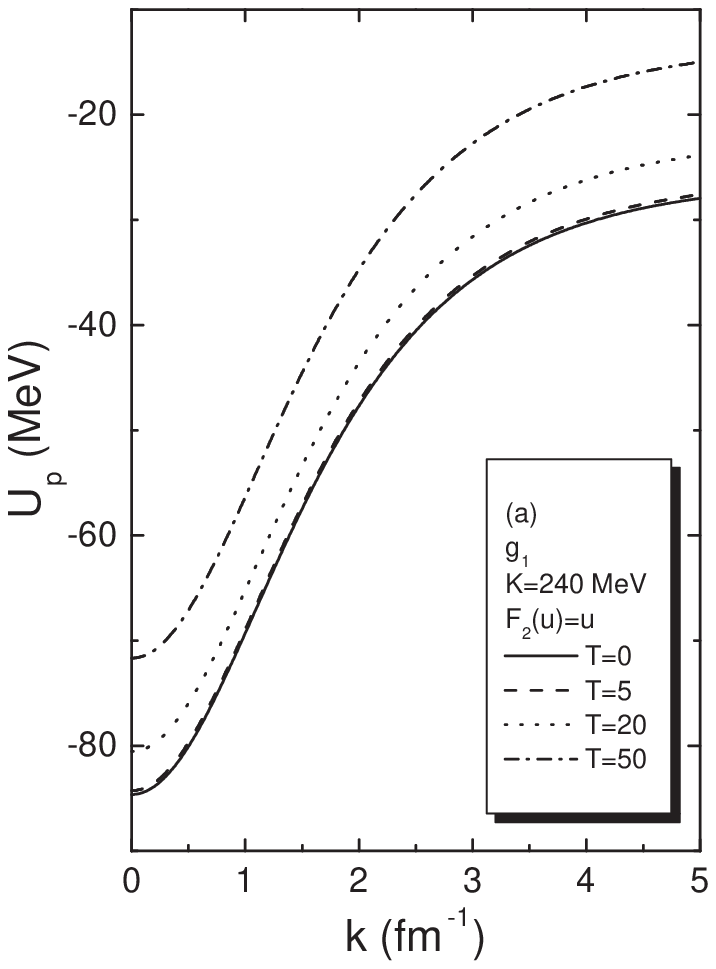}\
\includegraphics[height=8.0cm,width=8.0cm]{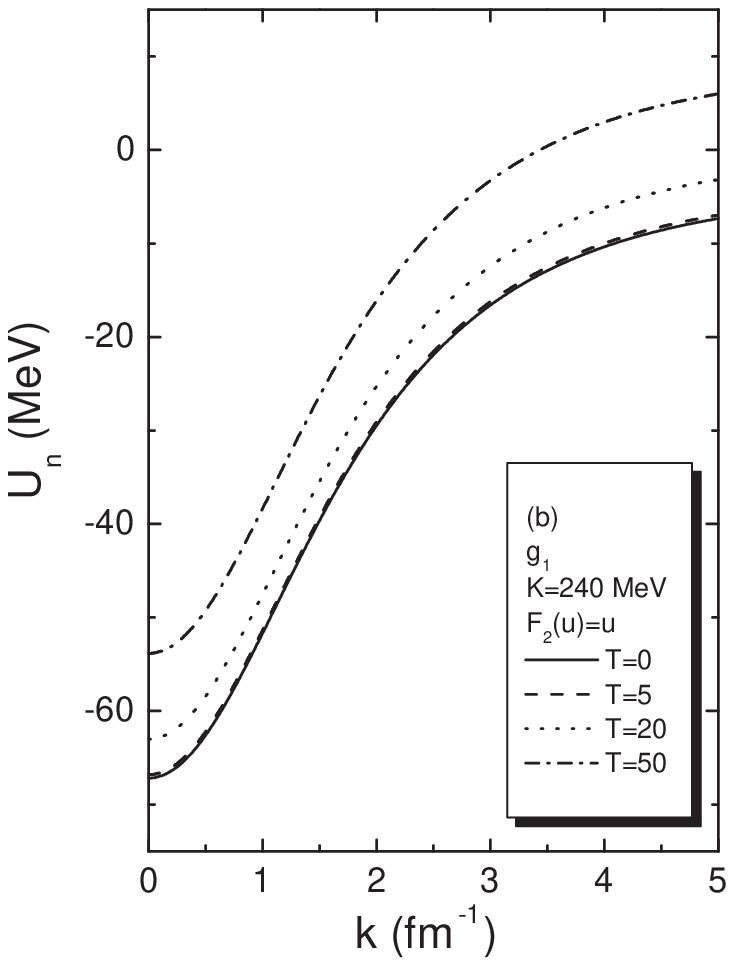}
\caption{$U_p$ and $U_n$, in case $g_1$ and for $K=240$ and
$F_2(u)=u$ as a function of temperature $T$ (in MeV) (at the
saturation density $n_0$). } \label{}
\end{figure}

\begin{figure}
\centering
\includegraphics[height=5.5cm,width=5.5cm]{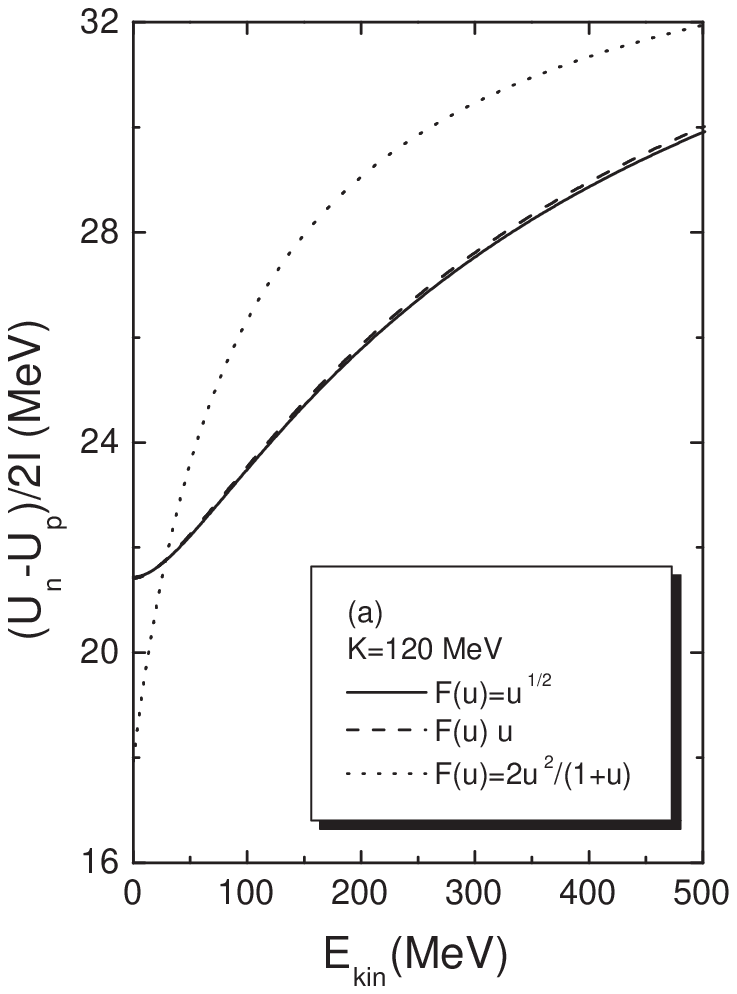}\
\includegraphics[height=5.5cm,width=5.5cm]{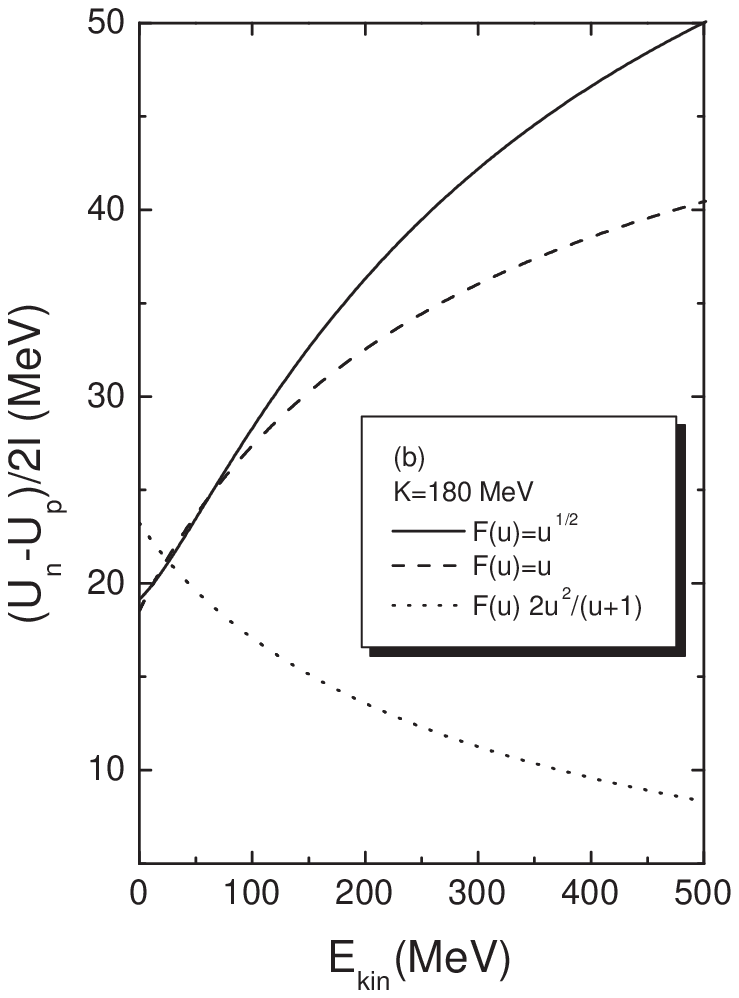}\
\includegraphics[height=5.5cm,width=5.5cm]{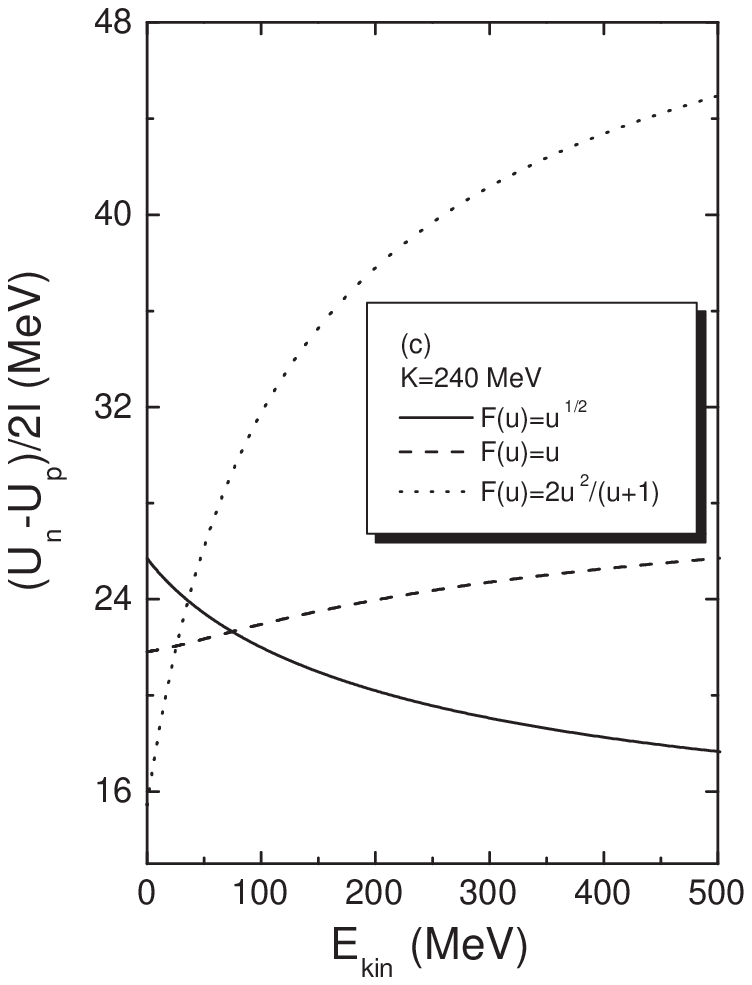}
\caption{The nuclear symmetry potential $U_{sym}=(U_n-U_p)/2I$
versus the nucleon kinetic energy $E_{kin}$ in case $g_1$, for
three values of $K$ and in each case for three different choices
of $F(u)$ (at the saturation density $n_0$). } \label{}
\end{figure}
\begin{figure}
\centering
\includegraphics[height=8.0cm,width=8.0cm]{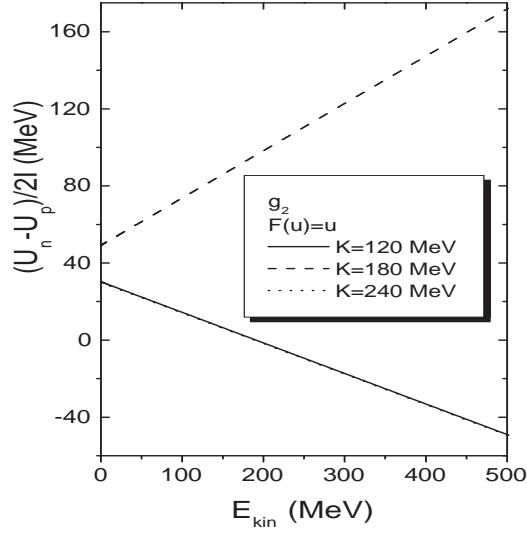}
\caption{$U_{sym}=(U_n-U_p)/2I$ versus $E_{kin}$ in case $g_2$,
for three values of $K$ (the cases $K=120$ MeV and $K=240$ MeV are
coincide) (at the saturation density $n_0$).} \label{}
\end{figure}
\begin{figure}
\centering
\includegraphics[height=8.0cm,width=8.0cm]{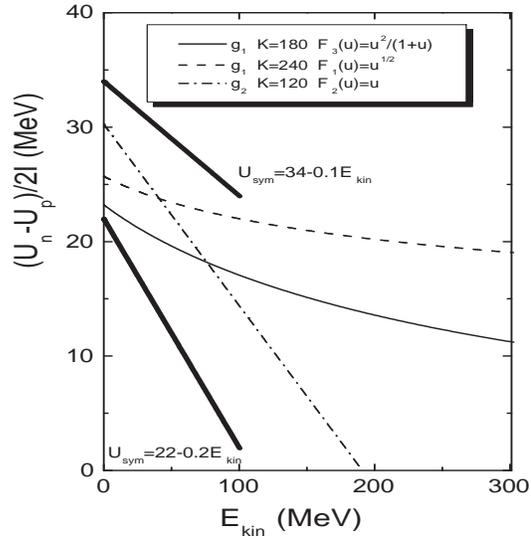}
\caption{$U_{sym}=(U_n-U_p)/2I$ versus $E_{kin}$ for the three
cases (at the saturation density $n_0$), which decreases with
increasing of nucleon momentum, in comparison with  $U_{sym}$
constrained by the experimental data. For more details see text.}
\label{}
\end{figure}
\begin{figure}
\centering
\includegraphics[height=8.0cm,width=8.0cm]{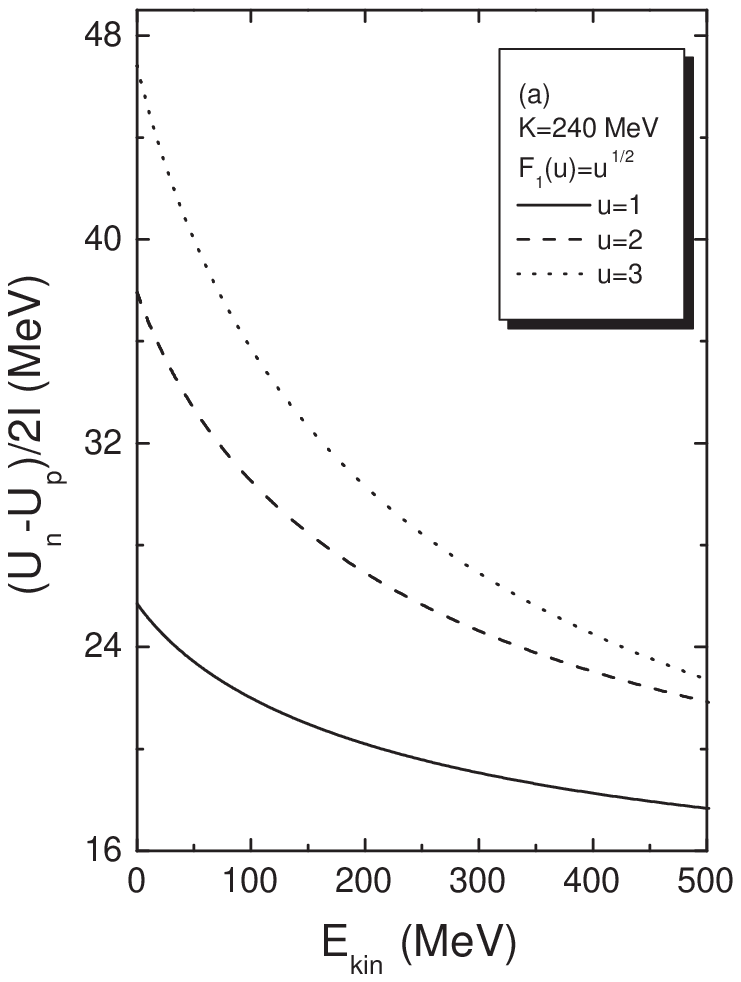}\
\includegraphics[height=8.0cm,width=8.0cm]{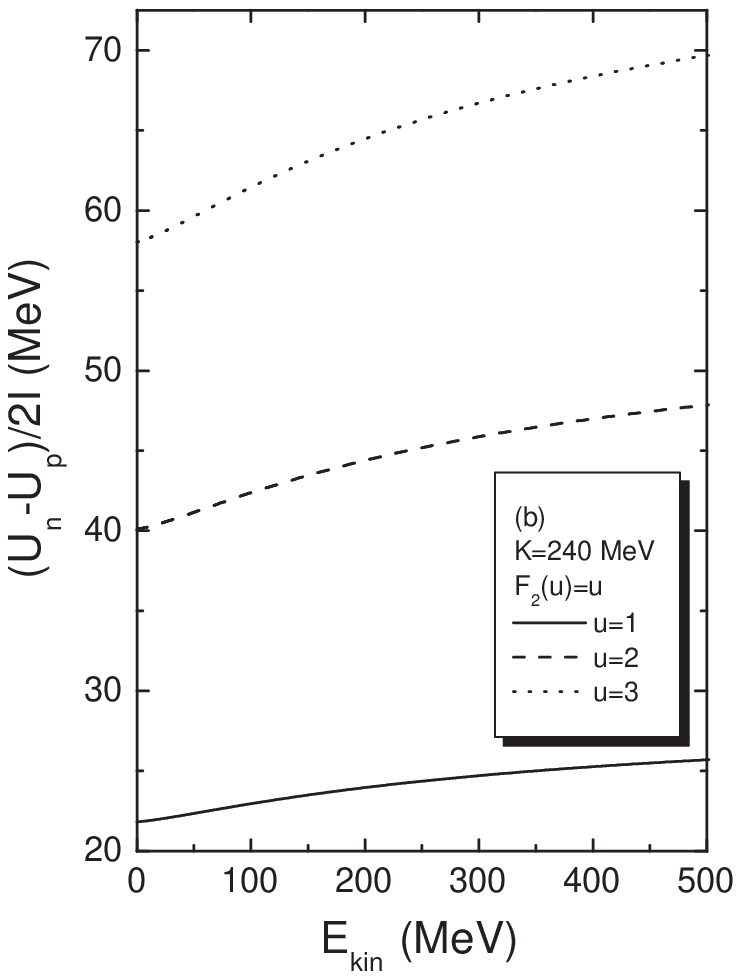}
\caption{$U_{sym}=(U_n-U_p)/2I$ versus $E_{kin}$ in case $g_1$,
for $K=240$ MeV  for three different values of $u$ (a) for
$F_1(u)=u^{1/2}$ and (b) for $F_2(u)=u$.} \label{}
\end{figure}
\begin{figure}
\centering
\includegraphics[height=8.0cm,width=8.0cm]{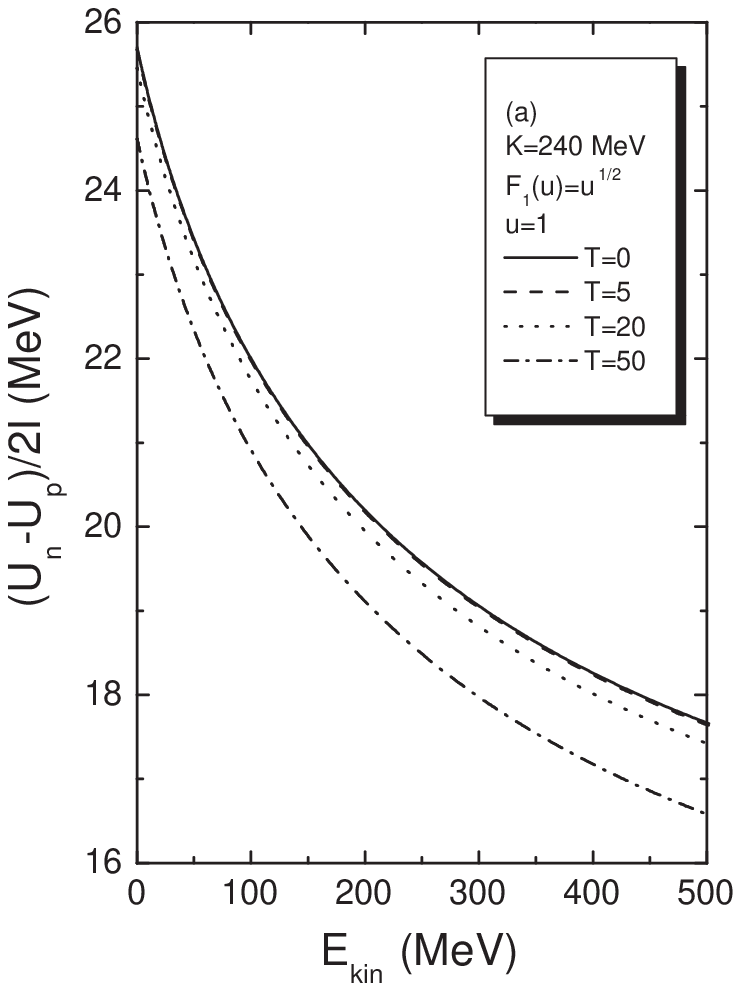}\
\includegraphics[height=8.0cm,width=8.0cm]{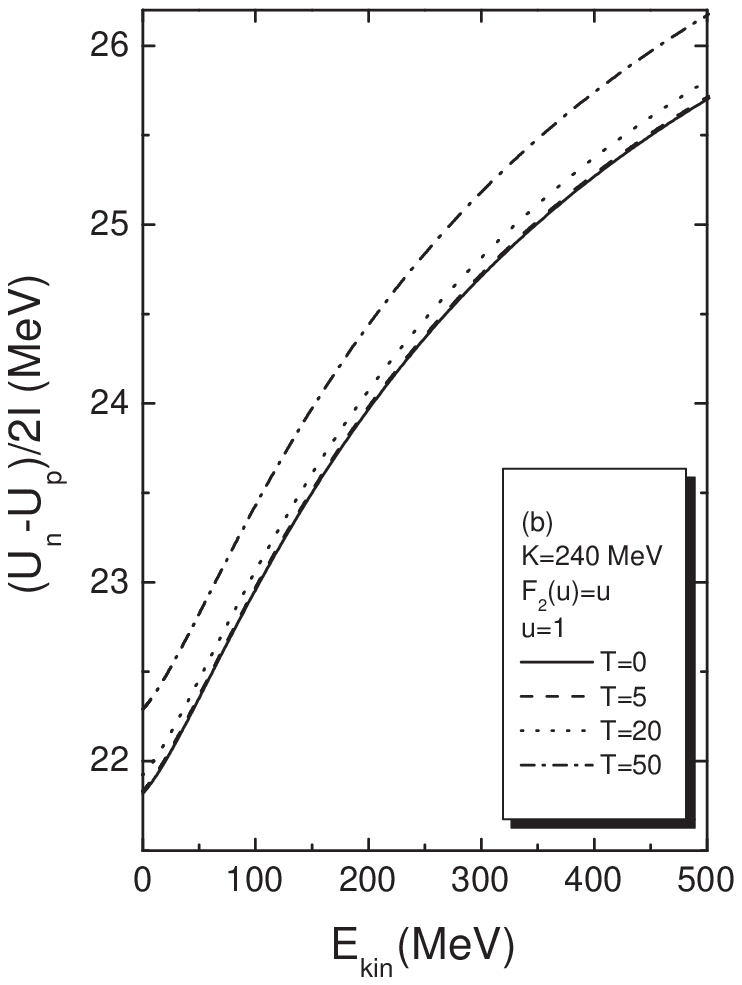}
\caption{$U_{sym}=(U_n-U_p)/2I$ versus $E_{kin}$ in case $g_1$,
for $K=240$ MeV  for various values of $T$ (in MeV) (a) for
$F_1(u)=u^{1/2}$ and (b) for $F_2(u)=u$. } \label{}
\end{figure}
\begin{figure}
\centering
\includegraphics[height=6.0cm,width=8cm]{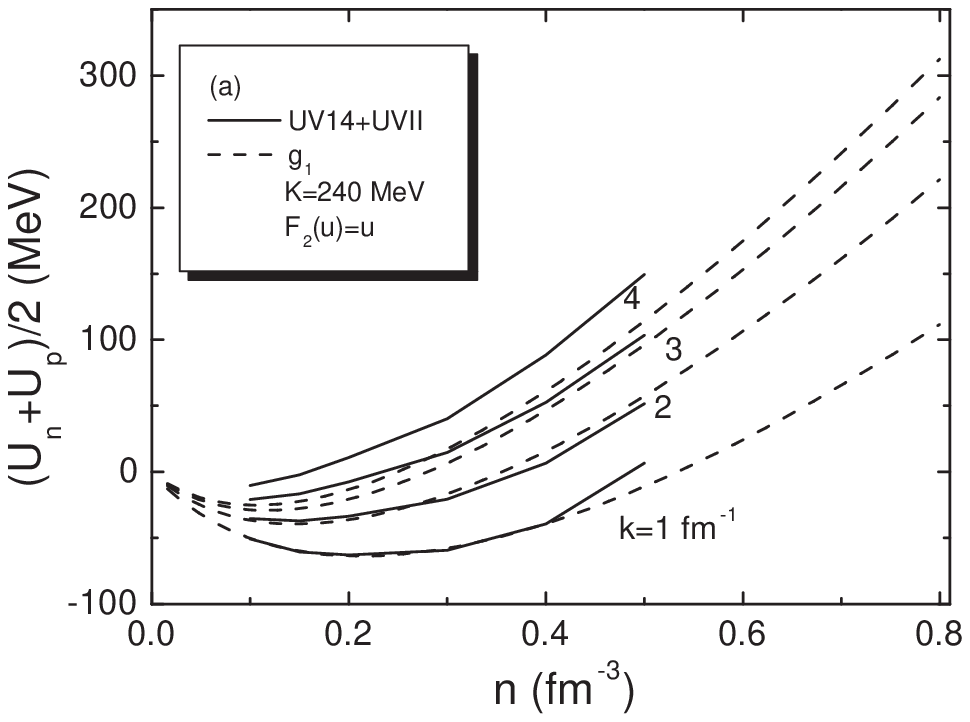}
\includegraphics[height=6.0cm,width=8cm]{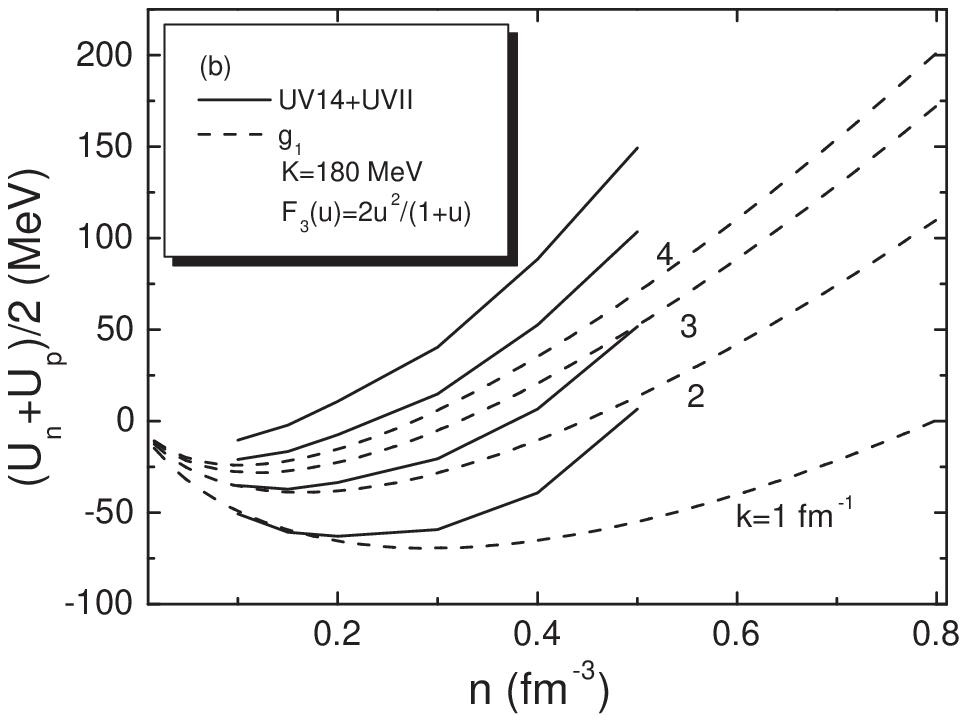}
\includegraphics[height=6.0cm,width=8cm]{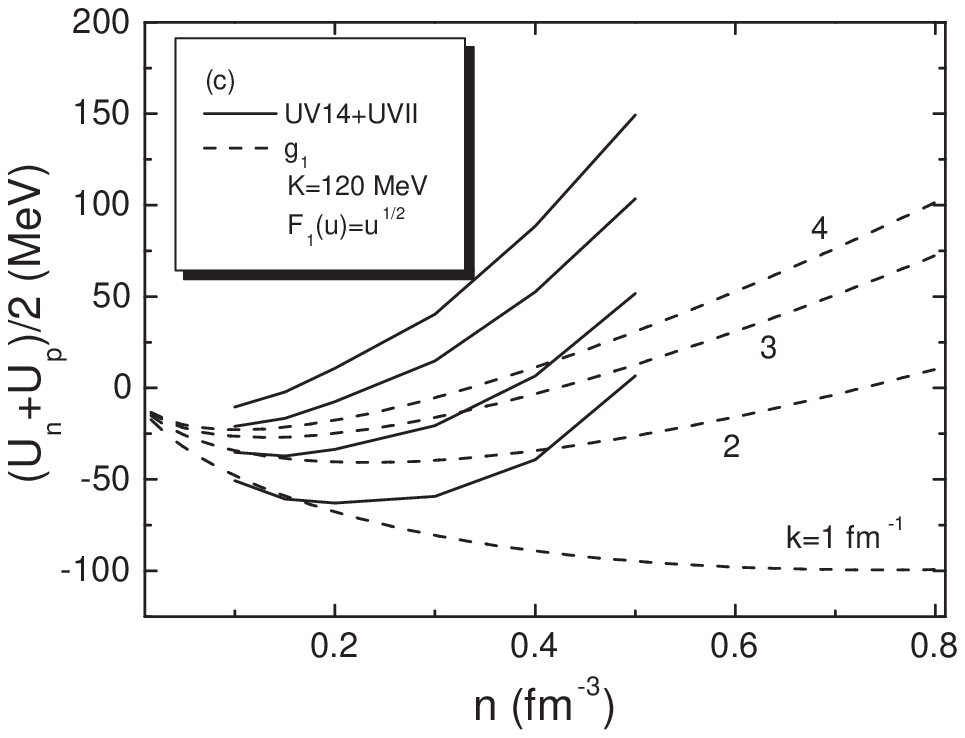}
\includegraphics[height=6.0cm,width=8cm]{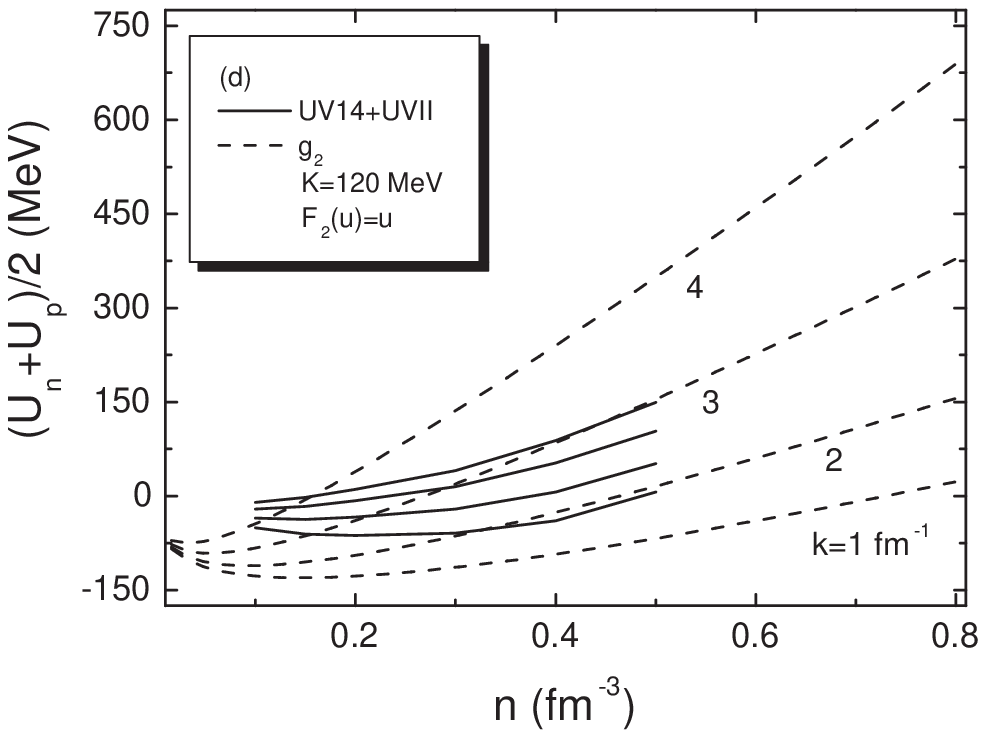}
\caption{The isoscalar potential $(U_n+U_p)/2$, as a function of
baryon density, at the four values of the momenta $k$, for four
different cases, in comparison with the variational many-body
calculations (UV14+UVII). } \label{}
\end{figure}
\begin{figure}
\centering
\includegraphics[height=5.5cm,width=5.5cm]{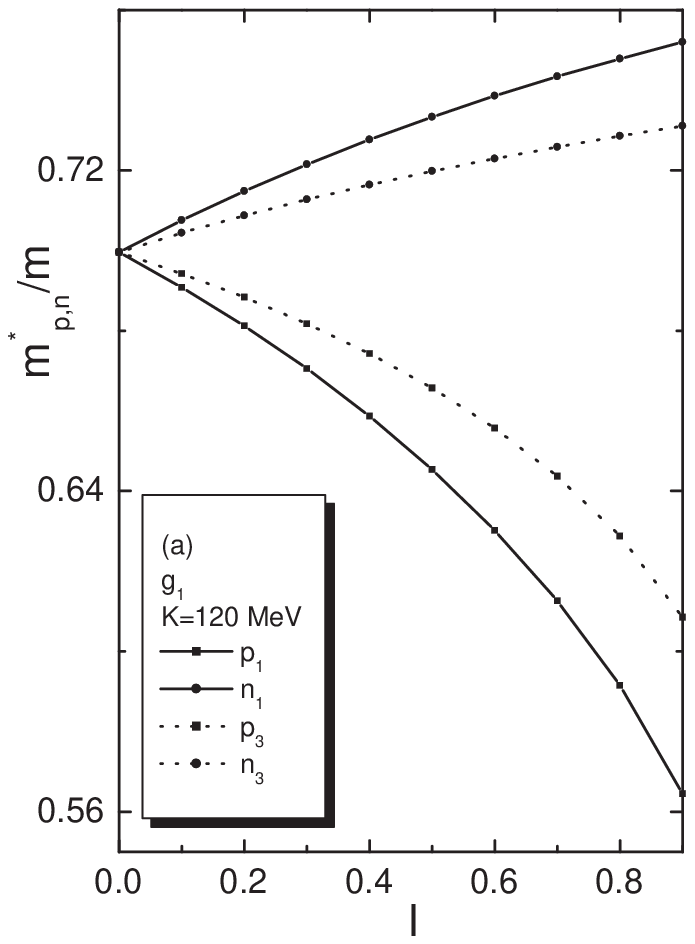}\
\includegraphics[height=5.7cm,width=5.7cm]{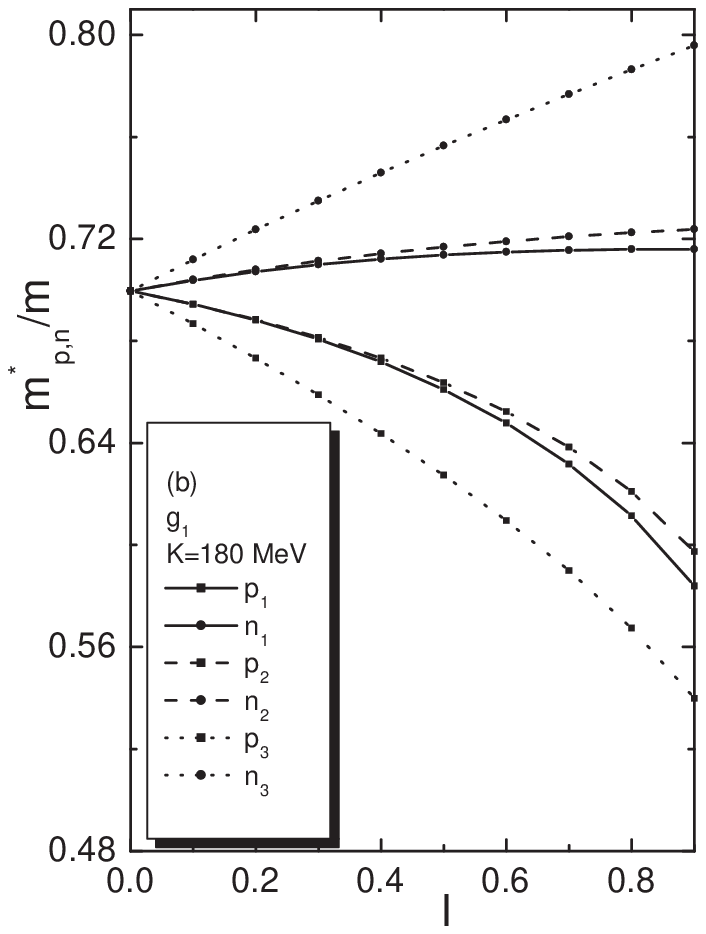}\
\includegraphics[height=5.5cm,width=5.5cm]{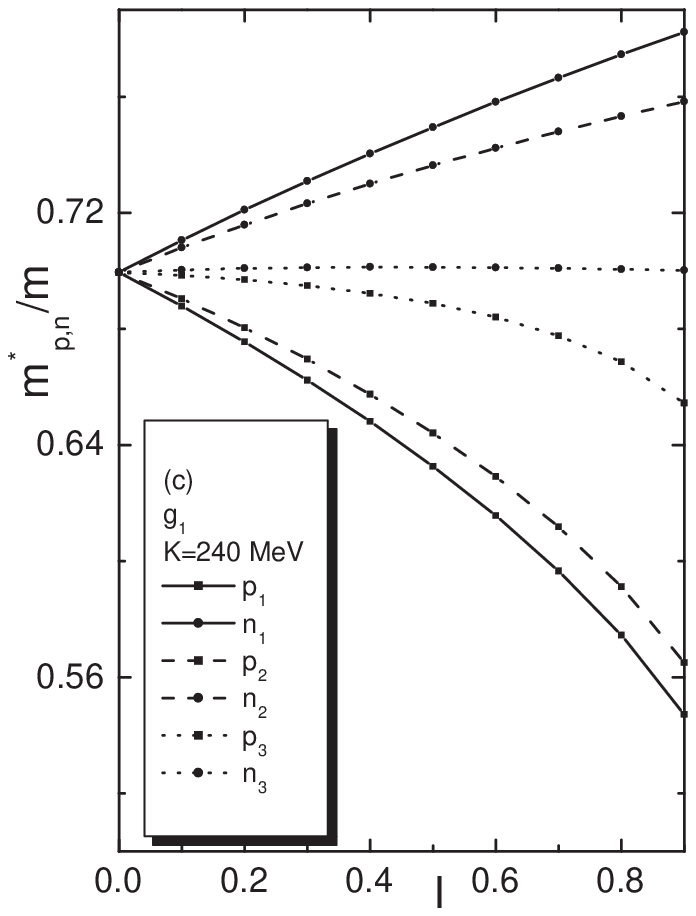}
\caption{The effective mass of proton and neutron as a function of
the asymmetry parameter $I$ for all the cases corresponding to
$g_1$ (at the saturation density $n_0$). } \label{}
\end{figure}
\begin{figure}
\centering
\includegraphics[height=5.5cm,width=5.5cm]{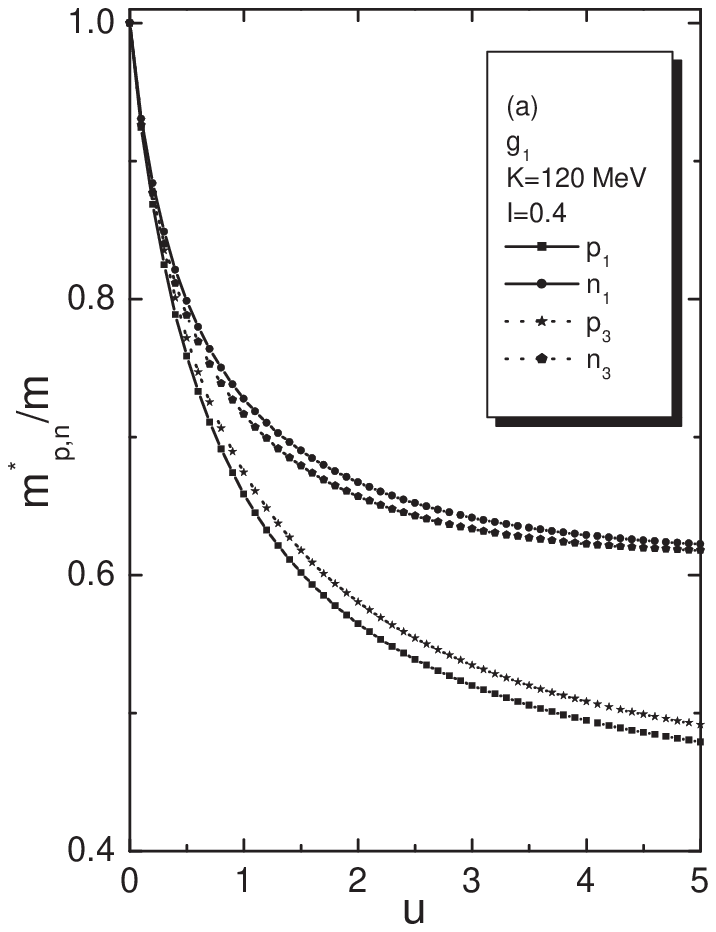}\
\includegraphics[height=5.5cm,width=5.5cm]{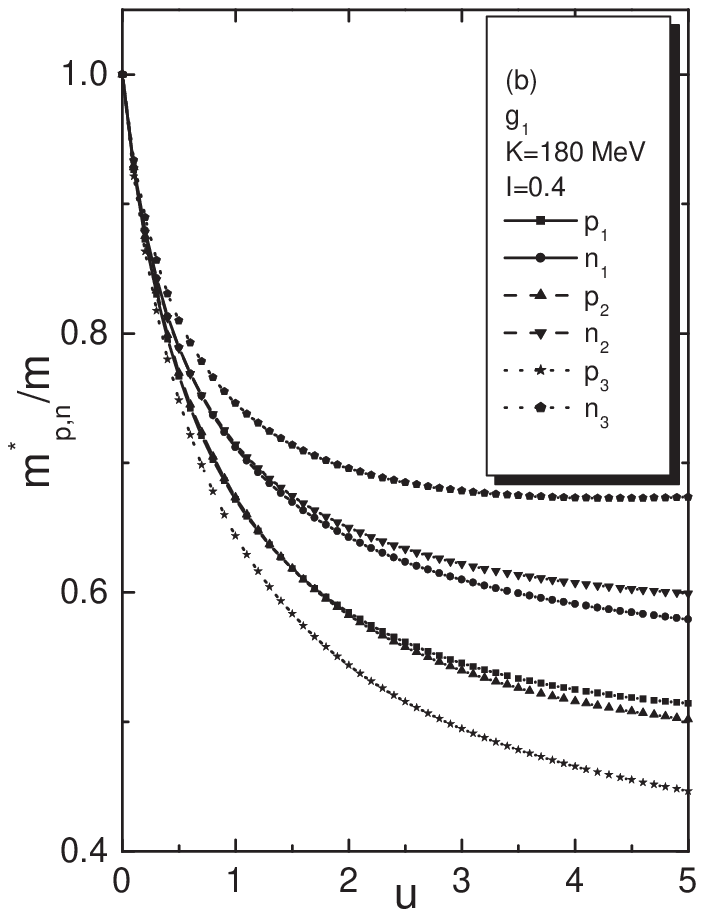}\
\includegraphics[height=5.5cm,width=5.5cm]{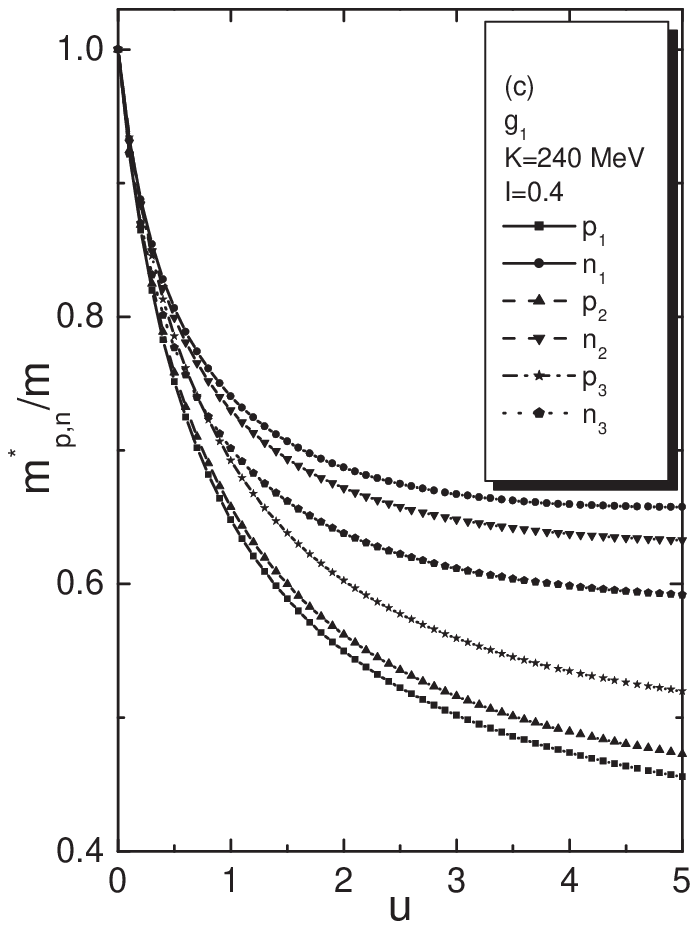}
\caption{The effective mass of proton and neutron versus $u$ for
all the cases corresponding to $g_1$. } \label{}
\end{figure}
\begin{figure}
\centering
\includegraphics[height=8.0cm,width=8.0cm]{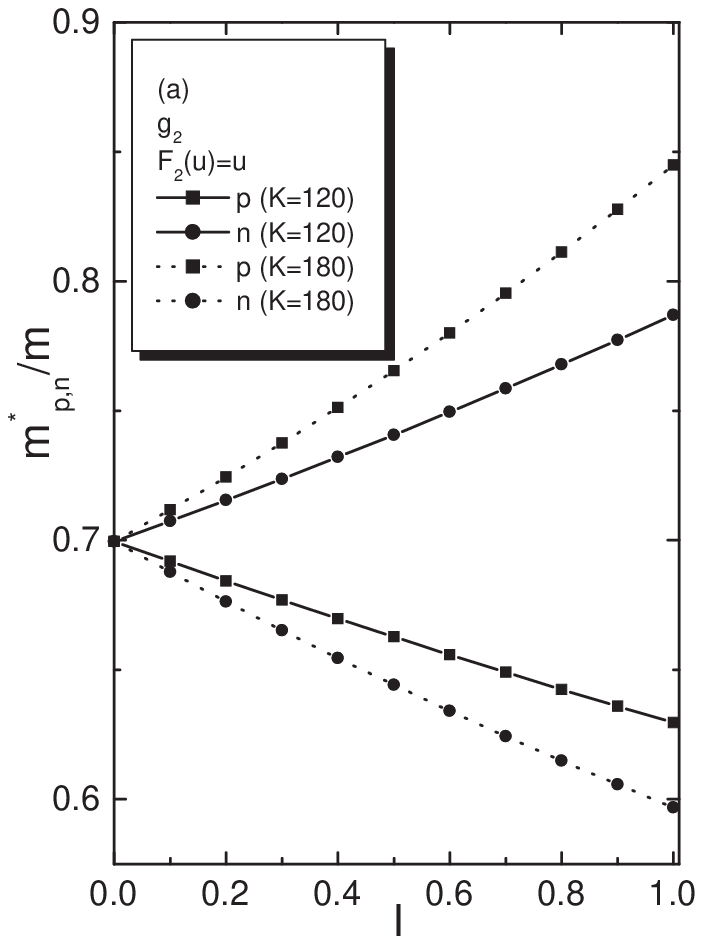}\
\includegraphics[height=8.0cm,width=8.0cm]{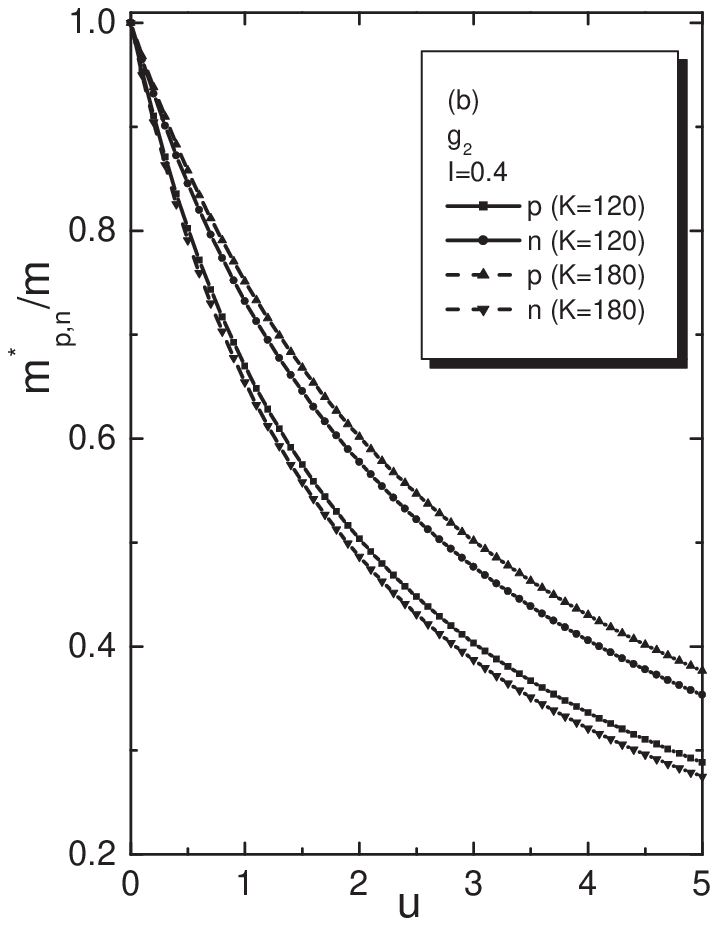}
\caption{(a) The effective masses of proton and neutron versus $I$
for two cases corresponding to $g_2$. (b) The effective masses of
proton and neutron versus $u$ for two cases corresponding to
$g_2$. } \label{}
\end{figure}


\end{document}